\documentclass[3p,authoryear,nopreprintline]{elsarticle}
\journal{Computers and Electronics in Agriculture}
\usepackage{graphicx}                    
\usepackage{hyperref}                    
\usepackage{lineno}
\usepackage{multirow}
\usepackage{multicol}
\usepackage{tikz}
\usepackage{xspace}
\usepackage{pgfplots}
\usepackage{subfig}
\usepackage{amsmath}
\usepackage{placeins}
\usepackage{float}
\usepackage{times}  
\usepackage{catchfile} 
\usetikzlibrary{positioning,calc,arrows.meta,shapes.misc}
\usepackage[capitalize]{cleveref}  

\newcommand{\titleOfArticle}{End-to-end pipeline for simultaneous temperature estimation and super resolution of low-cost uncooled infrared camera frames for precision agriculture applications}%
\newcommand{\frameLR}{I_\text{LR}\xspace}%
\newcommand{\frameSR}{I_\text{SR}\xspace}%
\newcommand{\frameNUC}{I_\text{NUC}\xspace}%
\newcommand{\frameGT}{I_\text{GT}\xspace}%
\newcommand{\scaleFactor}{\alpha\xspace}%
\newcommand{\nChannels}{\gamma\xspace}%
\newcommand{\nFrames}{\mathcal{N}\xspace}%
\newcommand{\tamb}{t_{amb}\xspace}%
\newcommand{\tobj}{t_{obj}\xspace}%
\newcommand{\taucamera}{\href{https://www.flir.com/products/tau-2/}{\texttt{Tau2}}\xspace}%
\newcommand{\blackbody}{\href{https://www.ci-systems.com/sr-800n-superior-accuracy-blackbody}{\texttt{SR-800N}}\xspace}%
\newcommand{\scientificCamera}{\href{https://www.flir.com/products/a655sc/}{\texttt{A655sc}}\xspace}%
\Crefname{figure}{Fig.}{Figures}
\crefname{section}{Sect.}{Sections}
\Crefname{table}{Table}{Tables}
\Crefname{equation}{Eq.}{Equations}
\begin{document}
\begin{frontmatter}
    \title{\titleOfArticle}%
    \author[1,2]{Navot Oz\corref{cor1}}
    \ead{navotoz@mail.tau.ac.il}
    \cortext[cor1]{Corresponding author}
    \author[3]{Nir Sochen}
    \author[2]{David Mendlovic}
    \author[1]{Iftach Klapp}
    \affiliation[1]{organization={Department of Sensing, Information and Mechanization Engineering, Agricultural Research Organization},
        addressline={Volcani Institute, P.O. Box 15159},
        postcode={7505101},
        city={Rishon LeZion},
        country={Israel},
    }
    \affiliation[2]{organization={School of Electrical Engineering, Tel Aviv University},
        addressline={P.O. Box 39040},
        postcode={6997801},
        city={Tel Aviv},
        country={Israel},
    }
    \affiliation[3]{organization={School of Mathematical Sciences, Tel Aviv University},
        addressline={P.O. Box 39040},
        postcode={6997801},
        city={Tel Aviv},
        country={Israel}
    }
    \begin{abstract}
        Radiometric infrared (IR) imaging is a valuable technique for remote-sensing applications in precision agriculture, such as irrigation monitoring, crop health assessment, and yield estimation.
        Low-cost uncooled non-radiometric IR cameras offer new implementations in agricultural monitoring.
        However, these cameras have inherent drawbacks that limit their usability, such as low spatial resolution, spatially variant nonuniformity, and lack of radiometric calibration.

        In this article, we present an end-to-end pipeline for temperature estimation and super resolution of frames captured by a low-cost uncooled IR camera.
        The pipeline consists of two main components: a deep-learning-based temperature-estimation module, and a deep-learning-based super-resolution module.
        The temperature-estimation module learns to map the raw gray level IR images to the corresponding temperature maps while also correcting for nonuniformity.
        The super-resolution module uses a deep-learning network to enhance the spatial resolution of the IR images by scale factors of $\times2$ and $\times4$.

        We evaluated the performance of the pipeline on both simulated and real-world agricultural datasets composing of roughly 20,000 frames of various crops.
        For the simulated data, the results were on par with the real-world data with sub-degree accuracy - $0.54^\circ C$ mean absolute error (MAE) for $\times 2$ scale factor, and $0.84^\circ C$ MAE for $\times 4$ scale factor.
        For the real data, the proposed pipeline was compared to a high-end radiometric thermal camera, and achieved sub-degree accuracy - $0.81^\circ C$ MAE for $\times 2$ scale factor, and $0.81^\circ C$ MAE for $\times 4$ scale factor. The results of the real data are on par with the simulated data.
        We show that our pipeline can compete with high-end thermal cameras in terms of quality and accuracy of the temperature and crop water stress index (CWSI) estimations using affordable hardware, with errors of $1.42\%$ for $\times2$ and $1.86\%$ for $\times4$ between the ground truth and the estimated CWSI\@.
        The runtime of the pipeline is less than $1_\text{sec}$ per frame on a CPU, allowing it to run at video rates.

        The proposed pipeline can enable various applications in precision agriculture that require high quality thermal information from low-cost IR cameras.
    \end{abstract}
    \nolinenumbers  
\end{frontmatter}

\section{Introduction}\label{sec:intro}
Radiometric infrared (IR) imaging is a valuable technique for remote sensing applications in precision agriculture.
It provides information about crop health~\citep{Mahlein2016,Caldern2014,Falkenberg2007,Alchanatis2005},
crop yield estimation~\citep{Das2020UAVThermalIA,Raeva2018,Pradawet2022} and soil moisture~\citep{BANERJEE2018,BERTALAN2022107262}.

IR imaging is well suited to irrigation monitoring.
\cite{irrigationGonzales2013} used IR imaging to assess the water status of trees using the crop water stress index (CWSI).
\cite{Osroosh2015} developed a novel method for CWSI determinations using IR imaging for irrigation scheduling.
\cite{CWSI_B} compared the results of irrigation scheduling using CWSI and IR imaging in vineyards.
\cite{CWSI} developed a protocol for precision irrigation using CWSI and IR imaging of peach trees.
Recently, \cite{Rozenfeld2024} used IR imaging to develop a spatial model for irrigation scheduling in orchards.

Long-wave infrared (LWIR) cameras measure the thermal radiation emitted by objects in the LWIR spectrum (usually $8-14_{\mu m}$).
These cameras have either a photon-counting (i.e., charge-coupled device (CCD)~\citep{infraredTheramlImaging}) or microbolometer-based sensor.
The CCD cameras are cryogenic cooled to reduce noise and increase sensor sensitivity.
On the other hand, the uncooled microbolometer-based IR cameras are simpler and thus less expensive, allowing for wider adoption in various applications.
The microbolometer sensor measures the change in resistance of a material due to the change in temperature inflicted by thermal radiance~\citep{bolometer}.

Uncooled microbolometer-based IR cameras can be broadly classified into two categories: radiometric and non-radiometric.

Radiometric IR cameras are calibrated so that their output can be interpreted as temperature.
These radiometric IR cameras are expensive because they require special calibration by both the user in the field and, periodically, the manufacturer. Moreover, they often require an internal shutter for one-point calibration of the camera, incurring complex control mechanizes and a surplus in price.

Non-radiometric IR cameras are not calibrated to measure temperature. They output gray-levels that are proportional to the intensity of the thermal radiation.

This work focuses on low-cost uncooled microbolometer-based non-radiometric IR cameras (hereafter referred to as uncooled IR cameras), bearing in mind their inherent drawbacks, which limit usability.
The first drawback is that the gray-level output of the uncooled IR camera needs to be calibrated to accurately estimate the temperature. The gray-level output also suffers from spatially variant nonuniformity, caused by: self-heating of the lens and camera onto the sensor itself, minute manufacturing variations in each microbolometer sensor, temperature drift caused by the ambient temperature of the camera, and various noises in the system~\citep{infraredTheramlImaging}.

The second drawback is that spatial resolution is low due to the trade-off between pixel size and sensitivity, which limits the ability to capture fine details.
As a comparison, a typical uncooled IR camera has tens of thousands of pixels, whereas a typical visible camera has millions of pixels.
The low spatial resolution of the uncooled IR cameras limits their effective range, because the camera is unable to capture fine details at a distance.
This is especially important in precision agriculture, where the camera is often mounted on an UAV, and is required to capture images of the crops in nadir from a distance. Due to the camera's low resolution, the UAV must be flown at lower altitudes, which increases the time and cost of the survey.
The resolution of the IR images can be increased by using super-resolution (SR) techniques, which upscale the low-resolution (LR) IR images to a higher resolution.
\newcommand{\EtoEwidthLR}{4em}
\newcommand{\EtoEwidthHR}{8em}
\newcommand{\EtoEfontSize}{\normalsize}
\newcommand{\EtoEdimensionsLR}{$\nFrames\times \text{h}\times \text{w}$}
\newcommand{\EtoENUCdimensionsLR}{$\text{1}\times \text{h}\times \text{w}$}
\newcommand{\EtoENUCdimensionsSR}{$\text{1}\times \pmb{\scaleFactor}\text{h}\times \pmb{\scaleFactor}\text{w}$}
\begin{figure*}[ht]
    \begin{tikzpicture}[auto, thick]
        \coordinate (start) at (2em,0);
        \coordinate (end) at ($(start) + (\linewidth-\EtoEwidthHR/3, 0)$);
        \node[anchor=center, align=center, rectangle] (lr) at (start) {\includegraphics[width=\EtoEwidthLR]{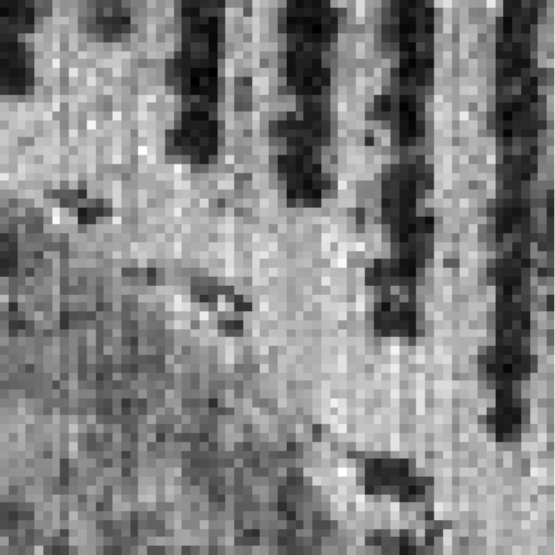}};
        \node[align=center, anchor=south, font=\EtoEfontSize]  at (lr.north) {$\frameLR$\\ \EtoEdimensionsLR};
        \node[align=center, anchor=north, font=\EtoEfontSize]  at (lr.south) {Gray levels};
        \node[anchor=east, align=right, rectangle] (sr) at (end) {\includegraphics[width={\EtoEwidthHR}]{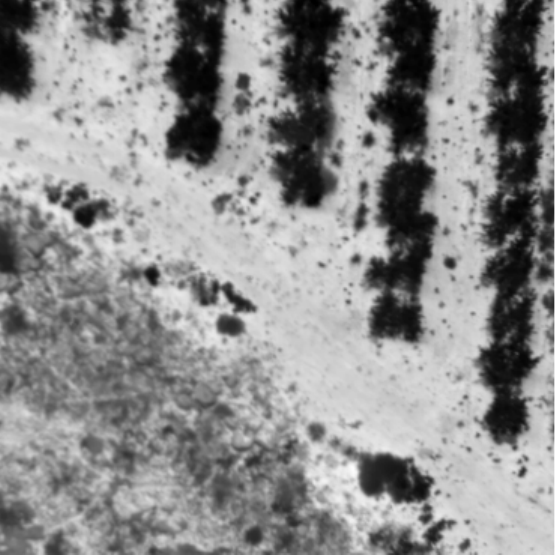}};
        \node[align=center, rectangle] (nuc) at ($(lr.east)!0.55!(sr.west)$) {\includegraphics[width=\EtoEwidthLR]{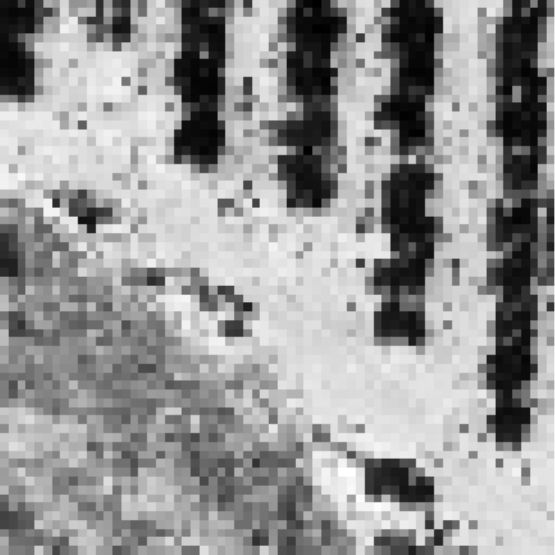}};
        \node[align=center, anchor=south, font=\EtoEfontSize]  at (nuc.north) {$\frameNUC$\\ \EtoENUCdimensionsLR};
        \node[align=center, anchor=north, font=\EtoEfontSize]  at (nuc.south) {Temperature};
        \node[draw, anchor=center, align=center, rectangle, text width=6.5em, minimum height=\EtoEwidthLR, font=\normalsize] (nucBlock) at ($(lr.east)!0.5!(nuc.west)$) {Nonuniformity\\ correction (NUC)};
        \node[align=center, anchor=south, font=\EtoEfontSize] (tamb)  at ($(nucBlock.north)+(0,0.8em)$) {Ambient temperature};
        \draw[-Latex] (tamb.south) -- (nucBlock.north);
        \node[draw, anchor=center, align=center, rectangle, text width=5em, minimum height=\EtoEwidthLR, font=\normalsize] (srBlock) at ($(nuc.east)!0.5!(sr.west)$) {Super\\ resolution (SR)};
        \node[align=center, anchor=south, font=\EtoEfontSize]  at (sr.north) {$\frameSR$\\ \EtoENUCdimensionsSR};
        \node[align=center, anchor=north, font=\EtoEfontSize]  at (sr.south) {Temperature};
        \draw[-Latex] (lr.east) -- (nucBlock.west);
        \draw[-Latex] (nucBlock.east) -- (nuc.west);
        \draw[-Latex] (nuc.east) -- (srBlock.west);
        \draw[-Latex] (srBlock.east) -- (sr.west);
    \end{tikzpicture}
    \caption{End-to-end pipeline for nonuniformity correction (NUC) followed by super resolution (SR).
    The two inputs to the pipeline are a low-resolution single-frame ($\nFrames=1$) or multiframe $\frameLR$ with dimensions $\nFrames\times h\times w$ and the ambient temperature $\tamb$ of the camera.
    First the scene temperature is estimated and its nonuniformity is corrected by the NUC module. The NUC module uses either a single frame (\cref{sec:methods:nuc:single}) or multiple frames (\cref{sec:methods:nuc:multi}). \cref{fig:methods:nuc} illustrates the NUC process.
    Second, the estimated temperature map $\frameNUC$ is super-resolved by the SR module (\cref{sec:methods:sr}) to obtain the output temperature estimation $\frameSR$ with dimensions $1\times\scaleFactor\cdot h\times\scaleFactor\cdot w$. \cref{fig:methods:sr} illustrates the SR process.}
    \label{fig:methods:e2e}
\end{figure*}
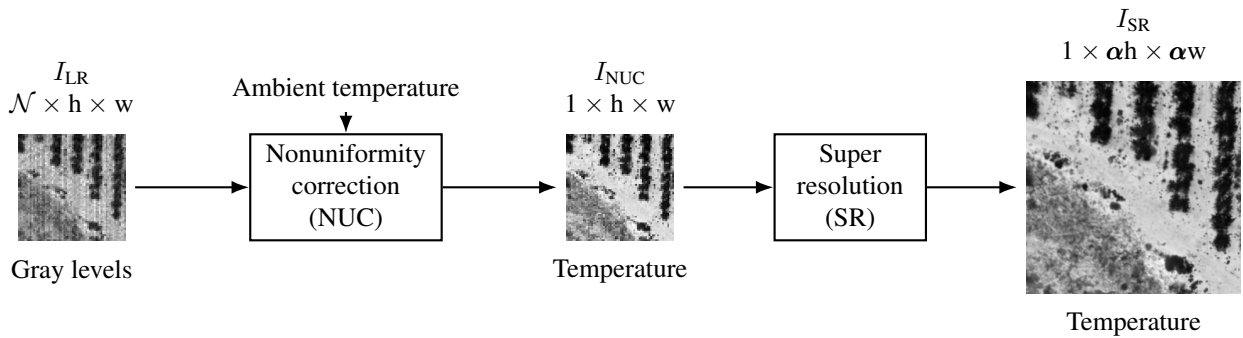
\subsection{Temperature estimation}\label{sec:intro:nuc}
As already noted, the gray level output of uncooled IR cameras is affected by the ambient temperature of the camera.
The physical relationship between the gray-level output and the scene temperature can be approximated by the Stefan-Boltzmann law.
In a small temperature environment, such as the earth's surface, the Stefan-Boltzmann law can be expended using a Taylor series.
The gray-level output of the uncooled IR camera is given by~\cite{Oz2023single}:

\begin{equation}\label{eq:cameraModel}
    L(\tobj, \tamb) = G(\tamb)\cdot \tobj +D(\tamb)
\end{equation} where $L(\tobj)$ is the gray-level output of the camera, $G(\tamb)$ is a gain term, $D(\tamb)$ is an offset term, and $\tamb$ and $\tobj$ are the ambient and scene temperatures, respectively.

There are several methods to estimate the temperature from the gray-level output of an uncooled IR camera.

The first group of methods aim to calibrate the output of the camera to known scene temperature. The simplest calibration method is to find a calibration coefficient for each pixel using the response curve of the camera, which is determined by measuring a range of scene temperatures.
The mathematical relations can be found using a one-point correction~\citep{Schulz1995}, a complex least-squares method~\citep{Nugent2013}, or polynomial fit of a known offset table~\citep{Liang2017}, where each of the methods relay on different assumptions and produce a different level of accuracy.
The underlying assumption in these methods is that the camera response curve is constant under different ambient temperatures.

The main disadvantage of the calibration methods is the need to calibrate for each individual camera separately for each ambient temperature and scene temperature. This requires collecting a large amount of calibrated data, which is time-consuming and expensive.
Moreover, the assumption of a constant camera response curve under different ambient temperatures is wrong, so the calibration coefficients do not hold for different ambient temperatures.

The second group of methods are scene-based, meaning that the temperature is estimated from the gray level output of the camera.
The scene-based methods can be broadly classified into two categories: single-frame and multiframe.

The single-frame methods estimate the temperature from a single frame, and they include histogram-based~\citep{admire}, neural network-based~\citep{nnOld}, and the more recent deep-learning-based~\citep{He2018,Jian2018,Chang2019,snrwdnn} methods.
\cite{Oz2023single} proposed a method to incorporate the physical model of the camera into the deep-learning model, and use the ambient temperature of the camera in the learning process.

The multiframe methods estimate the temperature from a sequence of frames.
They include shift calculations between consecutive frames~\citep{Harris99}, Kalman filtering~\citep{Averbuch2007}, and solving a complex inverse problem~\citep{Papini2018}.
Aerial surveys naturally collect sequences of frames with overlap. A major drawback of the multiframe methods is misalignment between the frames, which can cause significant errors in the temperature estimation.

To handle the misalignments in the frame sequences, \cite{Oz2023simultaneous} proposed using the redundant information between frames and incorporating the physical model and the ambient temperature into the model architecture.
This method is described in \cref{sec:methods:nuc:multi}.
\subsection{Super resolution}\label{sec:intro:sr}
The problem of estimating a high-resolution (HR) image from a LR image is tackled using SR methods, which include interpolation (e.g., nearest-neighbor, bicubic), sparse coding~\citep{sr_sparse_raw_im_patch,sr_via_sparse}, image priors~\citep{sr_grad_profile,fast_robust_sr} and example-based learning~\citep{glesner,sr_freeman}.
Recent SR methods have made use of deep learning~\citep{srcnn,vdsr,SRdense}.

\cite{Oz2020SR} proposed a method for fast and accurate SR of thermal images captured by uncooled IR cameras based on convolutional neural networks (CNNs) that uses~$3\%$ of the computational resources required for to the state-of-the-art methods.
\par In a previous work, \cite{Klapp2020} combined a nonuniformity correction (NUC) and SR in a single pipeline for uncooled IR cameras.
The proposed pipeline was based on solving an inverse problem for the NUC and on a deep-learning method for the SR\@.
Although the proposed method produced good results, it had several significant drawbacks which are addressed in the present work:
the NUC was based on the solver algorithm LSQR~\citep{LSQR}, and its runtime for a single image was on the order of hours, making it unfeasible for real-time applications.
Moreover, it was only tested on simulated data.
Deep learning has never been used to combine the NUC and SR for estimations of scene temperature in a single pipeline.

This work combines previous efforts into an end-to-end pipeline based on deep learning for the first time, to produce real-time high resolution thermal images from low-cost IR cameras in agriculture remote sensing.
The proposed pipeline first estimates the temperature from input gray-level frames captured by a low-cost uncooled IR camera, and then increases the spatial resolution of the estimated temperature map.
Compared to a scientific radiometric camera, the proposed method shows a sub-degree mean absolute error of $0.54^\circ C$ for $\times 2$ SR and $0.84^\circ C$ for $\times 4$ SR, demonstrating significant improvement over previous works. Moreover, the results show a very low error in CWSI estimation of about $1.5\%$ relative to CWSI resulted from the measurements of a scientific level radiometric camera.
A schematic illustration of the proposed end-to-end pipeline is shown in \cref{fig:methods:e2e}.
The contributions of this work are:
\begin{enumerate}
    \item An end-to-end deep-learning-based pipeline for combined temperature estimation and SR using uncooled IR cameras, producing radiometric-level thermography.
    \item Fast inference times of less than $1_\text{sec}$ per frame, running at video rates.
    \item The method is general and can be applied to any uncooled IR camera, without the need to recalibrate.
    \item The results are tested on real data collected in various agricultural fields and on CWSI estimations.
\end{enumerate}
\section{Materials and Methods}\label{sec:methods}
\newcommand{\yDistFromSep}{6em}
\newcommand{\heightTextBox}{4em}
\newcommand{\widthTextBox}{5em}
\newcommand{\widthImages}{4em}
\newcommand{\distFromFig}{0.01em}
\newcommand{\fontAboveBelowFigs}{\normalsize}
\newcommand{\fontBlock}{\Large}
\newcommand{\locNucBlock}{0.71}
\begin{figure*}
    \begin{tikzpicture}[node distance=auto, auto, thick]
        \coordinate (separator_start) at (2em,0);
        \coordinate (separator_end) at ($(separator_start) + (\linewidth-2em, 0)$);
        \draw [dashed, thick] (separator_start) to (separator_end) ;

        \node[draw, rectangle, align=center, font=\large, fill=white] (temperature_block) at ($(separator_start)!\locNucBlock!(separator_end) + (0, 0)$)  {$\tamb$ Ambient\\temperature};

        \node[align=center] (orig_image_multi_upper) at ($(separator_start) + (\widthImages / 2, -\yDistFromSep)$)  {\includegraphics[width=\widthImages, trim={{1ex} {1ex} {9ex} {9ex}}, clip=True]{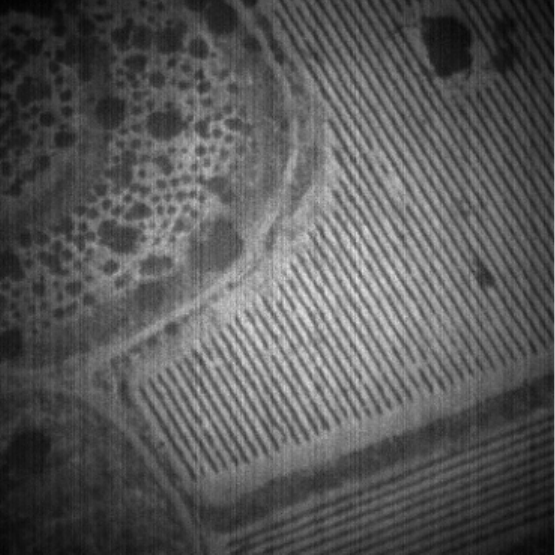}};
        \node[align=center]  at ($(orig_image_multi_upper) + (1ex, -1ex)$)  {\includegraphics[width=\widthImages,  angle=180, trim={{9ex} {9ex} {1ex} {1ex}}, clip=True]{figs/methods/NucMethods/input.pdf}};
        \node[align=center] (orig_image_multi) at ($(orig_image_multi_upper) + (2ex, -2ex)$)  {\includegraphics[width=\widthImages]{figs/methods/NucMethods/input.pdf}};
        \node[align=center]  at ($(orig_image_multi_upper) + (3ex, -3ex)$)  {\includegraphics[width=\widthImages, angle=-90, trim={{5ex} {5ex} {5ex} {5ex}}, clip=True]{figs/methods/NucMethods/input.pdf}};
        \node[align=center]  at ($(orig_image_multi_upper) + (4ex, -4ex)$)  {\includegraphics[width=\widthImages, angle=90, trim={{3ex} {2ex} {8ex} {8ex}}, clip=True]{figs/methods/NucMethods/input.pdf}};
        \node[above=\distFromFig+1ex of orig_image_multi, font=\fontAboveBelowFigs, align=center, rectangle] {$\nFrames$ Frames\\ gray levels};
        \node[align=center] (est_image_multi) at ($(separator_end) + (-\widthImages / 2, |-orig_image_multi)$)  {\includegraphics[width=\widthImages]{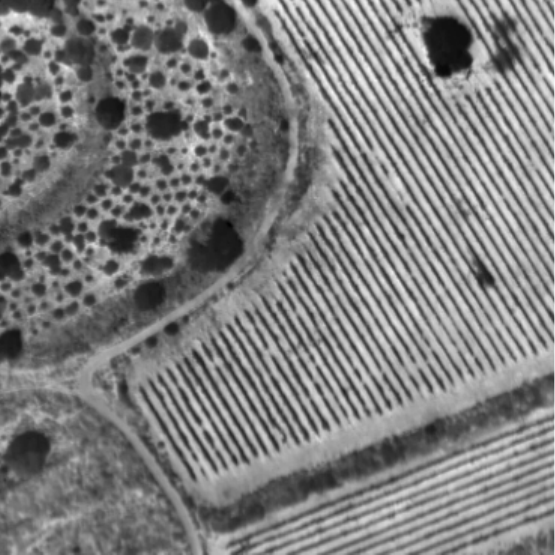}};
        \node[above=\distFromFig of est_image_multi, font=\fontAboveBelowFigs, align=center, rectangle] {Output\\ temperature};
        \node[draw, rectangle, minimum height=\heightTextBox, minimum width=\widthTextBox,  align=center, font=\fontBlock] (reg_multi_block) at ($(separator_start)!0.3!(separator_end) + (0, |-orig_image_multi)$)  {Registration};
        \node[align=center] (reg_image) at ($(separator_start)!0.51!(separator_end) + (0, |-orig_image_multi)$)  {\includegraphics[width=\widthImages]{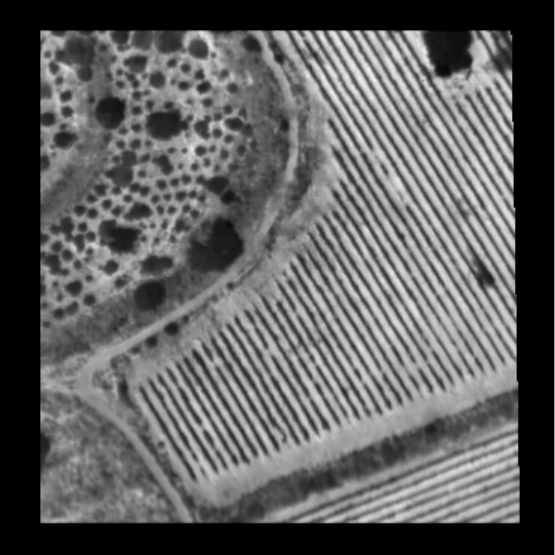}};
        \node[above=\distFromFig of reg_image, font=\fontAboveBelowFigs, align=center, rectangle] {Registered\\ frame};
        \node[draw, rectangle, minimum height=\heightTextBox, minimum width=\widthTextBox,  align=center, font=\fontBlock] (nuc_multi_block) at ($(separator_start)!\locNucBlock!(separator_end) + (0, |-orig_image_multi)$)  {NUC};
        \draw[-Latex] (orig_image_multi.east)  --  (reg_multi_block.west);
        \draw[-Latex] (reg_multi_block.east)  --  (reg_image.west);
        \draw[-Latex] (reg_image.east)  --  (nuc_multi_block.west);
        \draw[-Latex] (nuc_multi_block.east)  --  (est_image_multi.west);
        \draw[-Latex] (temperature_block.south)  --  (nuc_multi_block.north);

        \node[align=center] (orig_image_single) at ($(orig_image_multi|-, \yDistFromSep)$)  {\includegraphics[width=\widthImages]{figs/methods/NucMethods/input.pdf}};
        \node[below=\distFromFig of orig_image_single, font=\fontAboveBelowFigs, align=center, rectangle] {Frame\\ gray levels};
        \node[align=center] (est_image_single) at ($(separator_end) + (-\widthImages / 2, \yDistFromSep)$)  {\includegraphics[width=\widthImages]{figs/methods/NucMethods/output.pdf}};
        \node[below=\distFromFig of est_image_single, font=\fontAboveBelowFigs, align=center, rectangle] {Output\\ temperature};
        \node[draw, rectangle, minimum height=\heightTextBox, minimum width=\widthTextBox,  align=center, font=\fontBlock] (nuc_single_block) at ($(separator_start)!\locNucBlock!(separator_end) + (0, \yDistFromSep)$)  {NUC};
        \draw[-Latex] (orig_image_single.east)  --  (nuc_single_block.west);
        \draw[-Latex] (nuc_single_block.east)  --  (est_image_single.west);
        \draw[-Latex] (temperature_block.north)  --  (nuc_single_block.south);
    \end{tikzpicture}
    \caption{Schematic of the nonuniformity correction (NUC) methods.
        Top: single-frame NUC method~\citep{Oz2023single} described in \cref{sec:methods:nuc:single}.
        Bottom: multiframe NUC method~\citep{Oz2023simultaneous} described in \cref{sec:methods:nuc:multi}.
        Both methods require the ambient temperature of the camera, $\tamb$.
        The gray-level frame or frames are the input for the NUC method, and the estimated temperature map is the output.
        The dimension of all inputs and outputs are $h\times w$.}
    \label{fig:methods:nuc}
\end{figure*}
The end-to-end pipeline for temperature estimation and SR from low-cost uncooled IR cameras consists of two main stages~(\cref{fig:methods:e2e}).

The first stage, NUC, handles the spatial variations in pixel responses caused by factors such as the camera's ambient temperature, sensor noise, lens distortion, and various other noises.
Two NUC methods were used in this study: single-frame and multiframe.
The single-frame method utilizes a deep neural network to learn mapping from a gray-level frame to a temperature map.
The multiframe method exploits the temporal information from a sequence of overlapping gray-level frames to estimate the temperature map.
The output for both methods is the temperature map, $\frameNUC$.
Both methods are shown in \cref{fig:methods:nuc} and described in detail in \cref{sec:methods:nuc}.

The second stage, SR, increases the spatial resolution of the estimated temperature map $\frameNUC$.
The SR stage consists of a deep neural network that learns mapping from the LR temperature map to a HR temperature map.
The SR stage is shown in \cref{fig:methods:sr} and described in detail in \cref{sec:methods:sr}.
\subsection{Temperature estimation and NUC}\label{sec:methods:nuc}
Training supervised deep-learning methods requires large amounts of data~\citep{imagenet}.
Specifically, for the NUC problem, the required data are pairs of temperature maps and the corresponding gray-level frames, with pixel-perfect alignment.
In practice, it is extremely difficult and costly to collect such data, as it requires a scientific radiometric IR camera that is perfectly aligned to the low-cost IR camera, both collecting large amounts of data at different ambient temperatures. There are numerous challenges in this process (e.g., timing synchronization, alignment between cameras, varying ambient temperatures, etc.).

To overcome the data problem, a simulator was developed to simulate paired data for training in the NUC methods~\citep{Oz2023single}.
A low-cost IR camera (\taucamera) was placed in an environmental chamber. The chamber was placed in front of a scientific blackbody (\blackbody).
The environmental chamber was cyclically heated and cooled, and the \blackbody was also cycled through different temperatures.
The collected data were triplets ($\tamb$, $\tobj$, gray-level frames), where $\tamb$ is the ambient temperature, $\tobj$ is the temperature of the \blackbody, and the gray-level frames are the output of the \taucamera.

The simulator model was trained using these measurements. It was based on the linear properties of the camera response in a small temperature range, and on axis-symmetric pixel dependency.
The input to the simulator was an accurate temperature map and an ambient temperature, and the output was the simulated gray-level frame.
A thorough description of the simulator, including the dataset used to create it, can be found in \cite{Oz2023single}.

The simulator enabled the generation of a large number of frames with different temperatures and different nonuniformity patterns, to train the NUC methods in a supervised manner.

The training process of the end-to-end pipeline is shown in \cref{fig:methods:training}, along with the simulator.
\subsubsection{Single frame}\label{sec:methods:nuc:single}
To estimate the scene temperature from a single gray-level frame, the single-frame NUC method uses the camera's ambient temperature and a physical model of the camera. A schematic diagram of the method is shown in the top lane of \cref{fig:methods:nuc}.

The method develops a linear approximation of Stefan-Boltzmann's law, which links the scene temperature and the frame's gray level by a gain and an offset, both of which vary with the camera's ambient temperature.
The method constrains a neural network to learn the gain and offset terms, and incorporates the ambient temperature of the camera into the network itself.

The network was trained using the simulator from \cref{sec:methods:nuc}, which simulates gray-level IR frames from temperature maps and ambient temperatures. These gray-level frames are the input to the network. The original temperature maps used in the simulator were the ground truth (GT).

Detailed information about the network, its loss terms, the training method, and results on real data can be found in \cite{Oz2023single}.
\subsubsection{Multiple frame}\label{sec:methods:nuc:multi}
The multiframe NUC method exploits the redundant information found in multiple frames of the same scene to estimate the temperature of the scene.
There are two underlying assumptions in this method:
\begin{enumerate}
    \item The change in the ambient temperature of the camera across a short sequence of frames is negligible.
    \item There is overlap between frames, such that multiple views of the same scene are captured.
\end{enumerate}

The method is robust to misalignments between the frames, because rather than estimating the value of each pixel, it estimates a two-dimensional kernel around each pixel for every frame of the sequence.
The method learns to shift the center of gravity of the kernel so that it accounts for any misalignments between the frames.
For example, if the camera moves to the right between two frames, the method learns to shift the center of gravity of the kernel of the second frame to the right, so that the centers of the kernels for both pixels are aligned, even if the frames themselves are not.
The kernel is applied to each pixel and its surrounding pixels by multiplying it by the pixels and summing the results. 

The method utilizes the physical model of the camera to design a neural network that specifically estimates the average temperature of the scene from the camera's ambient temperature and gray-level output.
The estimated average temperature is then used as input to another neural network, to allow the second network to focus on the NUC, rather than estimating the offset between the gray-level output of the camera and the scene temperature.

To train the multiframe NUC module, the simulated frames from the simulator in \cref{sec:methods:nuc} were used as input and the original temperature map as the GT\@.
The simulated frames were randomly cropped, rotated and subjected to random perspective transformation to simulate the motion of the camera on a UAV\@.
The frames were then collected into bursts of $\nFrames$ frames, where $\nFrames$ is a hyperparameter of the method.
The frame bursts were then fed into a neural network which outputs a single estimated temperature map of the scene.
The method can also handle small registration misalignments between frames, which is a common problem in multiframe methods.
The bottom lane of \cref{fig:methods:nuc} shows a schematic diagram of the method.

A detailed explanation of the network, its loss terms, and the training method can be found in~\cite{Oz2023simultaneous}, who also demonstrated a performance increase with the number of frames in the burst.
\subsection{Super resolution}\label{sec:methods:sr}
\newcommand{\SRwidthLR}{4.5em}
\newcommand{\SRwidthHR}{8em}
\newcommand{\SRlocationOfPixelShuffle}{0.58}
\newcommand{\SRdistanceLaneFromMiddle}{3ex + \SRwidthLR / 2}
\newcommand{\SRdimensionsLR}{$\text{1}\times \text{h}\times \text{w}$}
\newcommand{\SRdimensionsFeatures}{$\pmb{\nChannels}\times \text{h}\times \text{w}$}
\newcommand{\SRdimensionsSR}{$\text{1}\times \pmb{\scaleFactor}\text{h}\times \pmb{\scaleFactor}\text{w}$}
\newcommand{\SRdimensionsFeaturesSR}{$\pmb{\nChannels}\times \pmb{\scaleFactor}\text{h}\times \pmb{\scaleFactor}\text{w}$}
\newcommand{\SRfontSize}{\normalsize}
\newcommand{\SRfontDimensions}{\normalsize}
\begin{figure*}
    \begin{tikzpicture}[auto, thick]
        \coordinate (start) at (2em,0);
        \coordinate (end) at ($(start) + (\linewidth-3*\SRwidthHR/7, 0)$);
        \node[anchor=center, align=center, rectangle] (lr) at (start) {\includegraphics[width=\SRwidthLR]{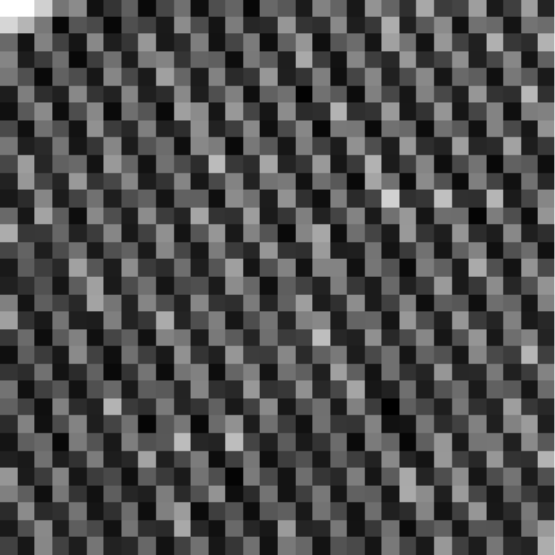}};
        \node[align=center, anchor=south, font=\SRfontSize]  at (lr.north) {$\frameNUC$\\ \SRdimensionsLR};
        \node[align=center, anchor=north, font=\SRfontSize] (lrLabel) at (lr.south) {Temperature};
        \node[anchor=east, align=right, rectangle] (sr) at (end) {\includegraphics[width={\SRwidthHR}]{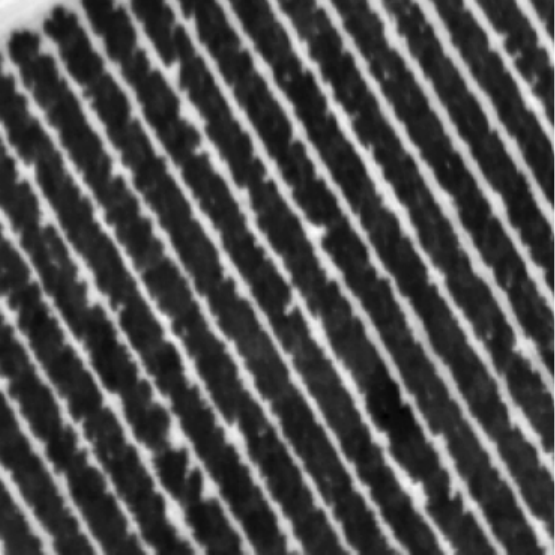}};
        \node[align=center, anchor=south, font=\SRfontSize]  at (sr.north) {$\frameSR$\\ \SRdimensionsSR};
        \node[align=center, anchor=north, font=\SRfontSize]  at (sr.south) {Temperature};
        \node[align=center, draw, rectangle, rotate=-90] (featureExtractor) at ($(lr.east)!0.08!(sr.west)$) {Feature extractor};
        \draw[-Latex] (lr.east) -- (featureExtractor.south);
        \coordinate (upperEast) at ($(featureExtractor.north) + (0, \SRdistanceLaneFromMiddle)$);
        \coordinate (upperWest) at ($(sr.west) + (0, |- upperEast)$);
        \coordinate (lowerEast) at ($(featureExtractor.north) - (0, \SRdistanceLaneFromMiddle)$);
        \coordinate (lowerWest) at ($(sr.west) + (0, |- lowerEast)$);
        \node[align=center, draw, rectangle] (processLR) at ($(upperEast)!0.24!(upperWest)$) {LR\\Processing};
        \draw[-Latex] (upperEast) -- (processLR.west);
        \node[align=center, draw, rectangle] (PixelShuffleLR) at ($(upperEast)!\SRlocationOfPixelShuffle!(upperWest)$) {Pixel\\ shuffle};
        \draw[-Latex] (processLR.east) -- (PixelShuffleLR.west);
        \node[align=center, draw, rectangle] (PixelShuffleEncoder) at ($(lowerEast)!\SRlocationOfPixelShuffle!(lowerWest)$) {Pixel\\ shuffle};
        \draw[-Latex] (lowerEast) -- (PixelShuffleEncoder.west);
        \node[draw, circle, inner sep=2pt] (concat) at ($(PixelShuffleEncoder.north)!0.5!(PixelShuffleLR.south)$) {C};
        \draw[-Latex] (PixelShuffleEncoder.north) -- (concat.south);
        \draw[-Latex] (PixelShuffleLR.south) -- (concat.north);
        \path let \p1 = ($(PixelShuffleEncoder.east)!0.5!(PixelShuffleLR.east)$) in \pgfextra{\xdef\endOfPixelShufflers{\x1}};
        \path let \p1 = (sr.west) in \pgfextra{\xdef\middleOfSR{\x1}};
        \node[align=center, draw, rectangle, rotate=-90] (srBlock) at ($(\endOfPixelShufflers,0)!0.3!(\middleOfSR,0 )$) {SR module};
        \draw[-Latex] (concat.east) -- (srBlock.south);
        \node[draw, circle, inner sep=2pt] (sumSR) at ($(srBlock.north)!0.5!(sr.west)$) {$+$};
        \draw[-Latex] (srBlock.north) -- (sumSR.west);
        \draw[-Latex] (sumSR.east) -- (sr.west);
        \newcommand{\lowestPntAddY}{1.4em}
        \path let \p1 = (lr) in \pgfextra{\xdef\xMiddleLr{\x1}};
        \path let \p1 = (PixelShuffleEncoder.south) in \pgfextra{\xdef\yLowLanePixelShuffle{\y1}};
        \coordinate (westBicArrowLowPnt) at ($(\xMiddleLr, \yLowLanePixelShuffle) - (0, \lowestPntAddY)$);
        \draw[-] (lrLabel.south) -- (westBicArrowLowPnt);
        \path let \p1 = (sumSR) in \pgfextra{\xdef\xMiddleSumSR{\x1}};
        \coordinate (eastBicArrowLowPnt) at ($(\xMiddleSumSR, \yLowLanePixelShuffle) - (0, \lowestPntAddY)$);
        \path let \p1 = (processLR) in \pgfextra{\xdef\xLowResBlock{\x1}};
        \path let \p1 = (westBicArrowLowPnt) in \pgfextra{\xdef\yBicArrow{\y1}};
        \node[align=center, draw, rectangle] (bicubicFrame) at ($(\xLowResBlock,0) + (0, \yBicArrow)$) {Bicubic Upscale};
        \draw[-Latex] (westBicArrowLowPnt) -- (bicubicFrame.west);
        \draw[-] (bicubicFrame.east) -- (eastBicArrowLowPnt);
        \draw[-Latex] (eastBicArrowLowPnt) -- (sumSR.south);
        \path let \p1 = ($(featureExtractor.north)!0.5!(processLR.west)$) in \pgfextra{\xdef\MiddleFeatureExtractorToProcessLR{\x1}};
        \node[rotate=-90, font=\SRfontDimensions] at (\MiddleFeatureExtractorToProcessLR, 0) {\SRdimensionsFeatures};
        \path let \p1 = ($(processLR.east)!0.5!(PixelShuffleEncoder.west)$) in \pgfextra{\xdef\MiddleProcessLRToPixelShuffle{\x1}};
        \node[rotate=-90, font=\SRfontDimensions] at (\MiddleProcessLRToPixelShuffle, 0) {\SRdimensionsFeatures};
        \path let \p1 = (srBlock.south) in \pgfextra{\xdef\MiddleWestSRBlock{\x1}};
        \path let \p1 = ($(\endOfPixelShufflers,0)!0.5!(\MiddleWestSRBlock,0)$) in \pgfextra{\xdef\MiddleEndOfPixelShuffleToSRBlock{\x1}};
        \node[rotate=-90, font=\SRfontDimensions] at (\MiddleEndOfPixelShuffleToSRBlock, 8ex) {\SRdimensionsFeaturesSR};
        \path let \p1 = (srBlock.north) in \pgfextra{\xdef\MiddleEastSRBlock{\x1}};
        \path let \p1 = ($(\middleOfSR,0)!0.5!(\MiddleEastSRBlock,0)$) in \pgfextra{\xdef\MiddleSrBlockToSR{\x1}};
        \node[rotate=-90, font=\SRfontDimensions] at (\MiddleSrBlockToSR, 9ex) {\SRdimensionsSR};
    \end{tikzpicture}
    \caption{The super-resolution (SR) method for IR frames first extracts features from $\frameLR$ using a convolutional neural network (CNN).
        The features are then processed in two lanes.
        The upper lane processes the features in the low-resolution (LR) space to save on computations, then upscales these features using the pixel-shuffle block by a factor of $\pmb{\scaleFactor}$.
        The lower lane immediately upscales the features using the pixel-shuffle block.
        After the pixel-shuffle blocks, the two lanes are combined by channel-wise concatenation and a CNN block is applied to create a single output channel (SR block in the figure).
        The output of the SR block is summed pixel-wise with the bicubic upscaled $\frameLR$ to create the final $\frameSR$.
        The dimensions of the feature maps are denoted as $\nChannels\times h\times w$, where $\nChannels$ is the number of channels, and $h$ and $w$ are the height and width of the feature maps, respectively.
        Notice that after the feature extraction, the dimensions of the feature maps remain constant until the pixel shuffler.
        The method is detailed in~\cite{Oz2020SR}.}
    \label{fig:methods:sr}
\end{figure*}
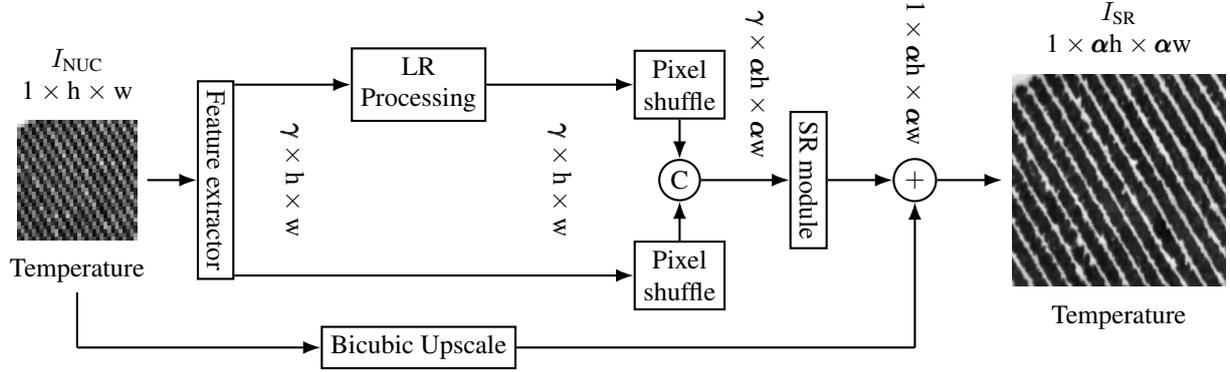
The SR method used in this work is based on the method presented in~\cite{Oz2020SR} and schematically in \cref{fig:methods:sr}. The method is based on a U-NET CNN~\citep{unet}.

The SR module, which utilizes the aforementioned SR method, is placed after the NUC module in the end-to-end pipeline, as seen in \cref{fig:methods:e2e}.
The output of the NUC module is a LR temperature map, $\frameNUC$, with the same dimensions as the input $\frameLR$, $h\times w$, and it is the input to the SR module.
The output of the SR module is a HR temperature map, $\frameSR$, with dimensions $\scaleFactor\cdot h\times \scaleFactor\cdot w$.
It exploits the low frequencies already present in the LR images, and uses the CNN to generate only the high frequencies.
The features of the input $\frameNUC$ are first extracted by a CNN layer.
These features are processed by a number of CNN residual blocks~\citep{Oz2020SR}, and then upscaled to HR by the pixel-shuffle layer~\citep{pixelShuffle}.
Parallel to the residual blocks, the features of $\frameNUC$ are also upscaled as is by a pixel-shuffle layer.
The two feature tensors are concatenated channel-wise, and processed by a final CNN block to create a single-channel output with spatial dimensions $1\times \scaleFactor\cdot h\times \scaleFactor\cdot w$.
The output of the final CNN block is summed pixel-wise with the bicubic upscaled $\frameNUC$ to create the final $\frameSR$.

The training process minimizes the difference between the HR temperature maps, $\frameGT$, and the $\frameSR$ generated by the network.
To create the input to the network during training, the HR temperature maps, $\frameGT$, are downscaled by a factor of $\scaleFactor$, creating the LR temperature maps, $\frameNUC$.
The output of the SR module is the upscaled temperature maps, $\frameSR$.
Precise details of the SR network architecture and training process are provided in~\cite{Oz2020SR}.
\subsection{Evaluation}\label{sec:methods:eval}
Several metrics were used to evaluate the proposed end-to-end pipeline.

\paragraph*{Mean absolute error (MAE)} measures the average absolute difference between the estimated and GT temperature maps pixel-wise. It is defined as:

\begin{equation}\label{eq:methods:eval:mae}
    \mathcal{L}_{MAE} = \frac{1}{H\cdot W}\sum_{i=1}^{H}\sum_{j=1}^{W} {\left| \frameSR[i,j] - \frameGT[i,j] \right|}
\end{equation}
where $\frameGT$ is the GT temperature map and $H$ and $W$ are the height and width of the temperature maps, respectively.

\paragraph*{Peak signal-to-noise ratio (PSNR)} is a widely used metric to measure quality in image-restoration tasks. It is defined as:

\begin{equation}
    \mathcal{L}_{PSNR} = 10 \cdot \log_{10}\left(\frac{\mathcal{X}^2}{\mathcal{L}_{MSE}\left(\frameSR, \frameGT\right)}\right)
\end{equation}
where $\mathcal{X}$ is the maximum possible pixel value of the temperature maps across all datasets, and $\mathcal{L}_{MSE}$ is the mean squared error.

\paragraph*{Structural similarity index measure (SSIM)} is used to evaluate the similarity between two images, and is highly correlated to human perception~\citep{ssimLoss2017}.
The complete equation, which can be found in~\cite{ssimLoss2017}. Herein, it is defined as $\mathcal{L}_{SSIM}$. The Python function used for the SSIM calculation is from the \texttt{kornia} v0.71 library.

\paragraph*{Earth Mover’s Distance (EMD)} also known as the Wasserstein distance, is a metric that quantifies the dissimilarity between two probability distributions. Specifically, it measures the minimal cost required to transform one distribution into another and is a robust measure in image-restoration tasks~\citep{earthMovers}. In the context of temperature maps, the EMD can be used to measure the dissimilarity between the estimated and GT temperature distributions. The EMD is defined as:

\begin{equation}
    \mathcal{L}_{EMD} = \frac{1}{H\cdot W}\sum_{i=1}^{H}\sum_{j=1}^{W} f_{ij} \cdot d(x_i, y_j)
\end{equation}
where $f_{ij}$ represents the flow (amount of “earth” moved) from bin $x_i$ to bin $y_j$, and $d(x_i, y_j)$ is the distance (cost) between bins $x_i$ and $y_j$, defined in this work as the L2 norm.

\paragraph*{Crop Water Stress Index (CWSI)} is a metric that provides insights into the water-stress levels experienced by crops.
It relates the actual plant temperature $T_{\text{plant}}$ to reference temperatures of a well-hydrated plant, $T_{\text{wet}}$, and a severely stressed plant, $T_{\text{dry}}$~\citep{CWSI}.
When a plant's surface temperature is closely aligned with the reference temperature for a well-hydrated plant, sufficient water availability for transpiration and no stress are indicated.
Conversely, a plant surface temperature that approaches the reference temperature for a severely stressed plant signals water stress. CWSI serves as a straightforward and valuable index for assessing plant water stress using thermal images of land surface temperatures~\citep{CWSI_A, CWSI_B}.
To calculate the CWSI, the following formula is used:

\begin{equation}
    \text{CWSI} = \frac{T_{\text{plant}} - T_{\text{wet}}}{T_{\text{dry}} - T_{\text{wet}}}
\end{equation}
where $T_{\text{plant}}$ represents the actual plant surface temperature, $T_{\text{wet}}$ is the reference temperature for a well-hydrated plant, and $T_{\text{dry}}$ is the reference temperature for a severely stressed plant.

$T_{\text{plant}}$ is defined as the lower (coolest) $33\%$ of the temperature distribution, $T_{\text{dry}}$ is defined as $\tamb+7$, and $T_{\text{wet}}$ is defined as the lower (coolest) $5\%$ of the temperature distribution of the pixels containing plants and canopy~\citep{CWSI}.
\subsection{End-to-end pipeline}\label{sec:methods:e2e}
\newlength{\TrainingWidthHR}
\TrainingWidthHR = 4em
\newlength{\TrainingWidthLR}
\TrainingWidthLR = 0.5\TrainingWidthHR
\newcommand{\TrainingDimensionsHR}{$\text{h}\times \text{w}$}
\newcommand{\TrainingDimensionsLR}{$\frac{1}{\scaleFactor}\text{h}\times \frac{1}{\scaleFactor}\text{w}$}
\newcommand{\TrainingLabelFontSize}{\normalsize}
\newcommand{\TrainingBlockFontSize}{\Large}
\begin{figure*}
    \begin{tikzpicture}[auto, thick]
        \coordinate (start) at (2em,0);
        \coordinate (end) at ($(start) + (\linewidth-3*\TrainingWidthHR/2, 0)$);
        \coordinate (heightUpperText) at (0, 2em + \TrainingWidthHR);
        \node[anchor=center, align=center, rectangle] (input) at (start) {\includegraphics[width=\TrainingWidthHR]{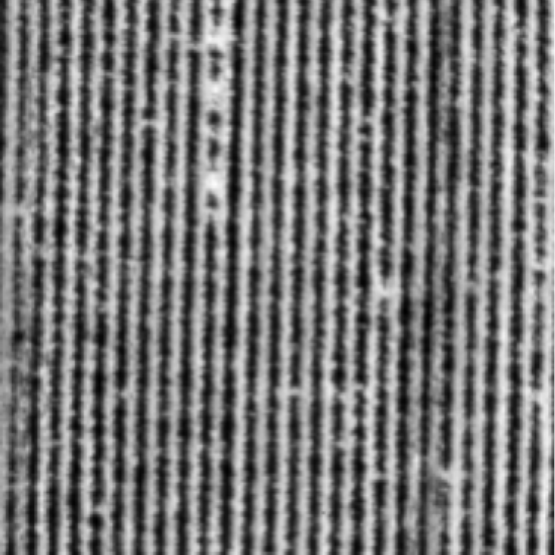}};
        \node[align=center, anchor=north, font=\TrainingLabelFontSize] (labelInput) at ($(input) + (0, |- heightUpperText)$) {Temperature};
        \node[align=center, anchor=south, font=\TrainingLabelFontSize] at ($(labelInput.south) - (0, 2em)$)  {$\frameGT$\\ \TrainingDimensionsHR};
        \node[align=center, draw, rectangle, rotate=-90, font=\TrainingBlockFontSize] (downscale) at ($(input.east)!0.07!(end)$) {Downscale};
        \draw[-Latex] (input.east) -- (downscale.south);
        \node[align=center, rectangle] (lr) at ($(start)!0.26!(end)$) {\includegraphics[width=\TrainingWidthLR]{figs/methods/training/hr.pdf}};
        \node[align=center, anchor=north, font=\TrainingLabelFontSize] (labelLR)  at ($(lr) + (0, |- heightUpperText)$) {Temperature};
        \node[align=center, anchor=south, font=\TrainingLabelFontSize]  at ($(labelLR.south) - (0, 2em)$) {\TrainingDimensionsLR};
        \draw[-Latex] (downscale.north) -- (lr.west);
        \node[align=center, draw, rectangle, rotate=-90, font=\TrainingBlockFontSize] (simulator) at ($(start)!0.5!(end)$) {Simulator};
        \draw[-Latex] (lr.east) -- (simulator.south);
        \path let \p1 = ($(simulator.west)!0.15!(simulator.east)$) in \pgfextra{\xdef\ySimTamb{\y1}};
        \path let \p1 = ($(lr.east)!0.5!(simulator.south)$) in \pgfextra{\xdef\xSimTamb{\x1}};
        \path let \p1 = (simulator.south) in \pgfextra{\xdef\xSim{\x1}};
        \node[align=center, draw, rectangle, font=\normalsize] (inputTambSim) at ($(\xSimTamb, \ySimTamb)$) {$\tamb$};
        \path let \p1 = (inputTambSim.west) in \pgfextra{\xdef\ySim{\y1}};
        \draw[-Latex] (inputTambSim.east) -- ($(\xSim, \ySim)$);
        \node[circle, draw, rectangle, minimum size=2em] (priorSpatialDependence) at ($(\xSimTamb, -\ySimTamb)$) {};
        \draw (priorSpatialDependence.center) circle (0.5em);
        \draw (priorSpatialDependence.center) circle (1em);
        \node[cross out, dashed, draw=black, minimum size=2em, rotate=0] at (priorSpatialDependence.center) {};
        \node[align=center, font=\tiny] (labelSpatialDep) at ($(priorSpatialDependence.south) - (0, 0.7em)$) {Spatial\\ dependence};
        \draw[-Latex] (priorSpatialDependence.east) -- ($(\xSim, -\ySim)$);
        \node[align=center, rectangle] (lr_input) at ($(start)!0.63!(end)$) {\includegraphics[width=\TrainingWidthLR]{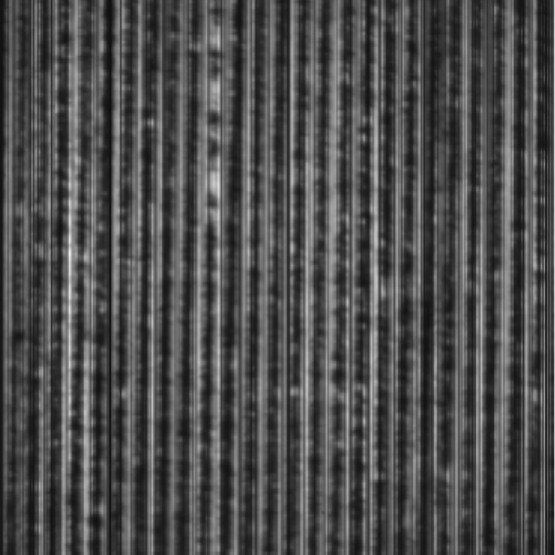}};
        \node[align=center, anchor=north, font=\TrainingLabelFontSize] (labelLRInput) at ($(lr_input) + (0, |- heightUpperText)$)  {Gray levels};
        \node[align=center, anchor=south, font=\TrainingLabelFontSize] at ($(labelLRInput.south)-(0, 2.5em)$) {$\frameLR$\\ \TrainingDimensionsLR};
        \draw[-Latex] (simulator.north) -- (lr_input.west);
        \node[align=center, draw, rectangle, rotate=-90, font=\TrainingBlockFontSize] (end2end) at ($(start)!0.8!(end)$) {End-to-End\\ pipeline};
        \draw[-Latex] (lr_input.east) -- (end2end.south);
        \node[align=center, rectangle] (output) at (end) {\includegraphics[width=\TrainingWidthHR]{figs/methods/training/hr.pdf}};
        \node[align=center, anchor=north, font=\TrainingLabelFontSize] (labelOutput) at ($(output) + (0, |- heightUpperText)$) {Temperature};
        \node[align=center, anchor=south, font=\TrainingLabelFontSize]   at ($(labelOutput.south)-(0,2em)$) {$\frameSR$\\ \TrainingDimensionsHR};
        \draw[-Latex] (end2end.north) -- (output.west);
        \newcommand{\lowestPntAddY}{1.4em}
        \path let \p1 = (labelSpatialDep.south) in \pgfextra{\xdef\yLowestPoint{\y1}};
        \path let \p1 = (input.center) in \pgfextra{\xdef\xWest{\x1}};
        \coordinate (westLossArrowLowPnt) at ($(\xWest, \yLowestPoint) - (0, \lowestPntAddY)$);
        \draw[-] (input.south) -- (westLossArrowLowPnt);
        \path let \p1 = (output.center) in \pgfextra{\xdef\xEast{\x1}};
        \coordinate (eastLossArrowLowPnt) at ($(\xEast, \yLowestPoint) - (0, \lowestPntAddY)$);
        \node[align=center, draw, rectangle, rotate=0, font=\TrainingBlockFontSize] (loss) at ($(westLossArrowLowPnt)!0.5!(eastLossArrowLowPnt)$) {Loss [\cref{eq:methods:loss}]};
        \draw[-Latex] (westLossArrowLowPnt) -- (loss.west);
        \draw[-Latex] (eastLossArrowLowPnt) --  (loss.east);
        \draw[-] (output.south) -- (eastLossArrowLowPnt);
    \end{tikzpicture}
    \caption{Training of the proposed pipeline.
        An accurate high-resolution temperature map $\frameGT$ with dimensions \TrainingDimensionsHR{}, is downscaled by $\scaleFactor$ to create a low-resolution (LR) temperature map with dimensions \TrainingDimensionsLR{}.
        The LR temperature map then enters the simulator along with a random $\tamb$ to create a gray-level frame with dimensions \TrainingDimensionsLR{}. The simulator utilizes the spatial dependence of the nonuniformity to create the gray-level frame.
        The gray-level frame $\frameLR$ is then used as input to the end-to-end pipeline from \cref{fig:methods:e2e}, which outputs a high-resolution temperature map estimation, $\frameSR$, with dimensions \TrainingDimensionsHR{}.
        The loss function compares the high-resolution temperature map estimation $\frameSR$ with the ground-truth $\frameGT$.}
    \label{fig:methods:training}
\end{figure*}
The two modules - temperature estimation and SR - were applied to the input $\frameLR$ to obtain the output $\frameSR$.
The entire pipeline is shown in \cref{fig:methods:e2e}.
The input $\frameLR$ with dimensions $h\times w$ and the ambient temperature of the camera $\tamb$ are processed by the NUC module.
Either the single-frame (\cref{sec:methods:nuc:single}) or multiframe (\cref{sec:methods:nuc:multi}) NUC method is used to estimate the temperature map $\frameNUC$ with the same dimensions as the input image.
Then, the estimated temperature map $\frameNUC$ is super-resolved (\cref{sec:methods:sr}) to obtain the output image $\frameSR$ with dimensions $\scaleFactor\cdot h\times\scaleFactor\cdot w$.

To train the entire pipeline, two steps were performed - pretraining and fine-tuning.
The pretraining step was done on each module separately. The single-frame NUC was trained as described in~\cite{Oz2023single}, the multiframe NUC as described in~\cite{Oz2023simultaneous}, and the SR as described in~\cite{Oz2020SR}. All weights were saved.

The fine-tuning training step was done by loading the pretrained weights and training the entire pipeline.
For each permutation - single-frame NUC and SR, multiframe NUC and SR - the weights of the pretrained modules were loaded and the entire pipeline was fine-tuned end-to-end.
The input to the pipeline was the LR frame and the output was the SR frame.
The new NUC and SR weights were saved separately.

The end-to-end pipeline was trained using two AdamW optimizers~\citep{adamw}, one for the NUC module and one for the SR module.
The initial learning rate of the SR module was set to $10^{-4}$ and that of the NUC module to $4\cdot10^{-5}$.
The learning rate was reduced by a factor of 0.5 after three consecutive epochs of plateaus in the validation loss (\cref{eq:methods:eval:mae}).
The batch size was set to 8 and the number of epochs was set to 60.
The loss function between the output $\frameSR$ and the $\frameGT$ was:
\begin{equation}\label{eq:methods:loss}
    \mathcal{L} = \mathcal{L}_{MAE}\left(\frameSR,\frameGT\right) + 10^{-3}\cdot\mathcal{L}_{SSIM}(\frameSR,\frameGT)
\end{equation} where $\mathcal{L}_{MAE}$ is the MAE and $\mathcal{L}_{SSIM}$ is the SSIM loss, both described in \cref{sec:methods:eval}.

The process of training the network along with the simulator is shown in \cref{fig:methods:training}.
The training convergence of the MAE and the learning rate is shown in \cref{appendix:fig:convergence} of the appendix.

All experiments were performed on an NVIDIA 4070Ti GPU using Python~3.11 and PyTorch~2.2.0.
\subsection{Data}\label{sec:methods:data}
\newcommand{\tableData}[1]{\begin{tabular}{|c|c|c|c|c|c|c|c|}
        \hline
        \multirow{2}*{Location} & \multirow{2}{*}{Date (yyyy-mm-dd)} & \multirow{2}{*}{Number of frames} & \multirow{2}{*}{Crop} & \multirow{2}{2.8cm}{\centering Ambient temperature [$^\circ C$]} & \multicolumn{2}{|c|}{Scene temperature [$^\circ C$]}       \\
        \cline{6-7 }
                                &                                    &                                   &                       &                                                                  & Min                                                  & Max \\
        \hline
        #1
    \end{tabular}}%
\begin{table*}[h!]
    \centering
    \caption{Statistics of the \textit{training} dataset.}
    \tableData{\hline
        Tzora & 2021-05-23 & 1,897 & Corn & 26.4 & 10.0 & 66.5\\
        \hline
        Barkan & 2019-08-12 & 2,211 & Cotton & 27.7 & 25.2 & 70.9\\
        \hline
        Gedera & 2019-08-11 & 538 & Cotton & 31.8 & 23.9 & 73.4\\
        \hline
        Neve Yaar & 2021-07-13 & 1,503 & Cotton & 35.2 & 12.5 & 70.9\\
        \hline
        Timurim & 2019-08-13 & 1,022 & Cotton & 34.5 & 27.6 & 71.1\\
        \hline
        Nir Eliyaho & 2021-10-05 & 2,477 & Peach & 28.4 & 10.1 & 70.3\\
        \hline
        Tzuba & 2018-08-05 & 675 & Peach & 31.8 & 27.7 & 80.7\\
        \hline
        Mevo Bytar & 2019-08-16 & 2,633 & Vines & 32.9 & 10.1 & 86.0\\
        \hline
        Mevo Bytar & 2021-08-25 & 765 & Vines & 31.2 & 10.1 & 84.9\\
        \hline
        Gilat & 2021-07-26 & 1,622 & Wheat & 33.8 & 10.0 & 79.1\\
        \hline
        Ramon & 2018-07-25 & 651 & Wheat & 40.3 & 27.4 & 89.1\\
        \hline
        \hline
        Total & 	 & 	 15,994 & 	  & 	  & 10.0 & 89.1 \\
        \hline}
    \label{tab:methods:data:train}
\end{table*}%
\begin{table*}[h!]
    \centering
    \caption{Statistics of the \textit{validation} dataset.}
    \tableData{\hline
        Barkan & 2019-08-12 & 542 & Cotton & 27.7 & 25.7 & 70.1\\
        \hline
        Timurim & 2019-08-13 & 277 & Cotton & 34.5 & 27.7 & 71.2\\
        \hline
        Nir Eliyaho & 2021-10-05 & 646 & Peach & 28.4 & 10.4 & 70.4\\
        \hline
        Tzuba & 2018-08-05 & 226 & Peach & 31.8 & 28.4 & 80.3\\
        \hline
        Mevo Bytar & 2019-08-16 & 1,487 & Vines & 32.9 & 12.8 & 89.9\\
        \hline
        Mevo Bytar & 2021-08-18 & 371 & Vines & 27.9 & 10.6 & 79.4\\
        \hline
        Gilat & 2021-08-09 & 360 & Wheat & 34.6 & 10.2 & 74.3\\
        \hline
        Ramon & 2018-07-25 & 217 & Wheat & 40.3 & 27.5 & 87.1\\
        \hline
        \hline
        Total & 	 & 	 4,126 & 	  & 	  & 10.2 & 89.9 \\
        \hline}
    \label{tab:methods:data:val}
\end{table*}%
The dataset used for training the network consisted of $15,994$ frames, each with dimensions of $640\times480$ pixels, captured from various agricultural fields across Israel.
The validation set, composed of $4,126$ frames, contained frames captured at the same locations at the same time, and frames of the same locations on different dates, to prevent data leakage and evaluate the network's ability to generalize to new data. The split between training and validation remained the same for the entire work.
\cref{tab:methods:data:train,tab:methods:data:val} present the statistics of the training and validation datasets, respectively.

The datasets were collected using an \scientificCamera camera mounted on a UAV at an altitude of $70-100_m$ above the ground.
The \scientificCamera accuracy was only $2\%$ of the temperature range in each frame.

The \scientificCamera was calibrated using FLIR ThermalResearch v2.1. The ambient temperatures used for calibration were collected using a mobile weather station at the locations during collection.

To create the nonuniformity simulator described in \cref{sec:methods:nuc}, an extensive dataset of calibration frames was captured using the \taucamera placed in an environmental chamber in front of a scientific-grade \blackbody blackbody. The \blackbody and environmental chamber were controlled to create sets of known temperatures, known ambient temperatures and gray-level frames. The frames were captured at various temperatures and ambient temperatures to create a dataset that covers a wide range of temperatures and nonuniformity patterns.

To simulate a general low-cost IR camera output, all built-in image processing by the \taucamera was disabled.
A single-point correction was done once using the camera built-in flat field calibration process. This correction can also be done using a lens cap in shutter-less cameras. More information about the calibration process can be found in~\cite{Oz2023single}.

Both the training and validation datasets are available upon request from the corresponding author. The calibration data are available at the end of the manuscript, in the data availability section.
\section{Results}\label{sec:results}
The following sections present the results of the proposed end-to-end pipeline.
\cref{sec:results:sim} discusses the results of the proposed pipeline on simulated data.
\cref{sec:results:real} presents results of the proposed pipeline on real data collected in the field.
\cref{sec:results:cwsi} shows the ability of the proposed pipeline to estimate the CWSI.
\subsection{Simulation}\label{sec:results:sim}
\begin{table*}
    \centering
    \caption{Comparison of metrics between the proposed methods single and multiframe nonuniformity correction (NUC) for both $\times2$ and $\times4$ scale factors.
        In each metric, the results in \textbf{bold} are the best result per scale factor.
        The multiframe results were obtained from 11 consecutive frames.
        MAE, mean absolute error; PSNR, peak signal-to-noise ratio; SSIM, structural similarity index measure; EMD, earth mover's distance}
    \scriptsize{
        \subfloat[Gilat, 2021-Aug-09]{\begin{tabular}{|c||c|c||c|c|}
                \hline
                Scale factor     & \multicolumn{2}{|c||}{2} & \multicolumn{2}{|c|}{4}                           \\
                \hline
                Type             & Single                   & Multi                   & Single & Multi          \\
                \hline
                \hline
                MAE [$^\circ C$] & 0.51                     & 0.51                    & 0.81   & \textbf{0.73}  \\ \hline
                PSNR [dB]        & 41.05                    & \textbf{41.09}          & 36.98  & \textbf{37.92} \\ \hline
                SSIM             & 0.95                     & \textbf{0.96}           & 0.88   & \textbf{0.89}  \\ \hline
                EMD [$^\circ C$] & 0.12                     & \textbf{0.09}           & 0.25   & \textbf{0.24}  \\ \hline
            \end{tabular}}
        \hfill\subfloat[Mevo Bytar, 2019-Aug-16]{\begin{tabular}{|c||c|c||c|c|}
                \hline
                Scale factor     & \multicolumn{2}{|c||}{2} & \multicolumn{2}{|c|}{4}                           \\
                \hline
                Type             & Single                   & Multi                   & Single & Multi          \\
                \hline
                \hline
                MAE [$^\circ C$] & \textbf{0.50}            & 0.52                    & 0.84   & \textbf{0.75}  \\ \hline
                PSNR [dB]        & \textbf{40.90}           & 40.56                   & 36.36  & \textbf{37.29} \\ \hline
                SSIM             & 0.96                     & 0.96                    & 0.90   & \textbf{0.91}  \\ \hline
                EMD [$^\circ C$] & \textbf{0.09}            & 0.10                    & 0.18   & \textbf{0.17}  \\ \hline
            \end{tabular}}
        \hfill\subfloat[Mevo Bytar, 2021-Aug-18]{\begin{tabular}{|c||c|c||c|c|}
                \hline
                Scale factor     & \multicolumn{2}{|c||}{2} & \multicolumn{2}{|c|}{4}                                  \\
                \hline
                Type             & Single                   & Multi                   & Single        & Multi          \\
                \hline
                \hline
                MAE [$^\circ C$] & \textbf{0.63}            & 0.64                    & 1.06          & \textbf{0.96}  \\ \hline
                PSNR [dB]        & \textbf{38.96}           & 38.62                   & 34.17         & \textbf{34.98} \\ \hline
                SSIM             & 0.95                     & \textbf{0.96}           & 0.88          & \textbf{0.89}  \\ \hline
                EMD [$^\circ C$] & \textbf{0.09}            & 0.14                    & \textbf{0.19} & 0.20           \\ \hline
            \end{tabular}}
        \hfill\subfloat[Nir Eliyaho, 2021-Oct-05]{\begin{tabular}{|c||c|c||c|c|}
                \hline
                Scale factor     & \multicolumn{2}{|c||}{2} & \multicolumn{2}{|c|}{4}                           \\
                \hline
                Type             & Single                   & Multi                   & Single & Multi          \\
                \hline
                \hline
                MAE [$^\circ C$] & 0.44                     & 0.44                    & 0.72   & \textbf{0.62}  \\ \hline
                PSNR [dB]        & \textbf{41.82}           & 41.60                   & 37.20  & \textbf{38.28} \\ \hline
                SSIM             & 0.97                     & 0.97                    & 0.93   & \textbf{0.94}  \\ \hline
                EMD [$^\circ C$] & \textbf{0.07}            & 0.09                    & 0.13   & 0.13           \\ \hline
            \end{tabular}}
        \hfill\subfloat[Ramon, 2018-Jul-25]{\begin{tabular}{|c||c|c||c|c|}
                \hline
                Scale factor     & \multicolumn{2}{|c||}{2} & \multicolumn{2}{|c|}{4}                           \\
                \hline
                Type             & Single                   & Multi                   & Single & Multi          \\
                \hline
                \hline
                MAE [$^\circ C$] & \textbf{0.31}            & 0.33                    & 0.45   & \textbf{0.38}  \\ \hline
                PSNR [dB]        & \textbf{44.98}           & 44.69                   & 41.79  & \textbf{43.11} \\ \hline
                SSIM             & 0.98                     & 0.98                    & 0.96   & \textbf{0.97}  \\ \hline
                EMD [$^\circ C$] & \textbf{0.08}            & 0.13                    & 0.13   & 0.13           \\ \hline
            \end{tabular}}
        \hfill\subfloat[Tzuba, 2018-Aug-05]{\begin{tabular}{|c||c|c||c|c|}
                \hline
                Scale factor     & \multicolumn{2}{|c||}{2} & \multicolumn{2}{|c|}{4}                           \\
                \hline
                Type             & Single                   & Multi                   & Single & Multi          \\
                \hline
                \hline
                MAE [$^\circ C$] & \textbf{0.58}            & 0.62                    & 1.08   & \textbf{0.99}  \\ \hline
                PSNR [dB]        & \textbf{39.37}           & 38.61                   & 33.40  & \textbf{34.10} \\ \hline
                SSIM             & 0.96                     & \textbf{0.97}           & 0.89   & \textbf{0.90}  \\ \hline
                EMD [$^\circ C$] & \textbf{0.09}            & 0.18                    & 0.19   & 0.19           \\ \hline
            \end{tabular}}
        \hfill\subfloat[Barkan, 2019-Aug-12]{\begin{tabular}{|c||c|c||c|c|}
                \hline
                Scale factor     & \multicolumn{2}{|c||}{2} & \multicolumn{2}{|c|}{4}                                  \\
                \hline
                Type             & Single                   & Multi                   & Single        & Multi          \\
                \hline
                \hline
                MAE [$^\circ C$] & \textbf{0.64}            & 0.75                    & 1.19          & \textbf{1.11}  \\ \hline
                PSNR [dB]        & \textbf{38.80}           & 37.39                   & 33.31         & \textbf{33.92} \\ \hline
                SSIM             & \textbf{0.97}            & 0.96                    & 0.89          & \textbf{0.90}  \\ \hline
                EMD [$^\circ C$] & \textbf{0.08}            & 0.17                    & \textbf{0.22} & 0.23           \\ \hline
            \end{tabular}}
        \hfill\subfloat[Timurim, 2019-Aug-13]{\begin{tabular}{|c||c|c||c|c|}
                \hline
                Scale factor     & \multicolumn{2}{|c||}{2} & \multicolumn{2}{|c|}{4}                           \\
                \hline
                Type             & Single                   & Multi                   & Single & Multi          \\
                \hline
                \hline
                MAE [$^\circ C$] & \textbf{0.67}            & 0.74                    & 1.26   & \textbf{1.17}  \\ \hline
                PSNR [dB]        & \textbf{38.55}           & 37.54                   & 32.84  & \textbf{33.49} \\ \hline
                SSIM             & 0.96                     & 0.96                    & 0.86   & \textbf{0.88}  \\ \hline
                EMD [$^\circ C$] & \textbf{0.11}            & 0.17                    & 0.33   & 0.33           \\ \hline
            \end{tabular}}
        \hfill\subfloat[Average]{\begin{tabular}{|c||c|c||c|c|}
                \hline
                Scale factor     & \multicolumn{2}{|c||}{2} & \multicolumn{2}{|c|}{4}                           \\
                \hline
                Type             & Single                   & Multi                   & Single & Multi          \\
                \hline
                \hline
                MAE [$^\circ C$] & \textbf{0.54}            & 0.57                    & 0.93   & \textbf{0.84}  \\ \hline
                PSNR [dB]        & \textbf{40.55}           & 40.01                   & 35.76  & \textbf{36.64} \\ \hline
                SSIM             & 0.96                     & 0.96                    & 0.90   & \textbf{0.91}  \\ \hline
                EMD [$^\circ C$] & \textbf{0.09}            & 0.13                    & 0.20   & 0.20           \\ \hline
            \end{tabular}\label{table:results:sim:avg}}}
    \label{table:results:sim}
\end{table*}
\subsubsection{Metric comparison}
\cref{table:results:sim} shows the validation dataset results of the proposed methods for single- and multiframe NUC for both scale factors ($\times2$ and $\times4$).
The results were obtained by averaging the metrics over each dataset.

The metrics in bold indicate the best results for each scale factor.
As seen in the dataset tables, as well as in the average \cref{table:results:sim:avg}, the multiframe NUC provided better results for $\times4$ while the single-frame NUC provided better results for $\times2$.
For $\times2$, the single-frame NUC MAE was better by $0.03^\circ \text{C}$, while for $\times4$, the multiframe NUC MAE was better by $0.09^\circ$C.

\newcommand{\captionResultsSimPatches}{Results of the end-to-end pipeline.
    All results were obtained using the multiframe nonuniformity correction (NUC) with 11 frames.
    GT, ground truth; SR, super resolution.}%
\newcommand{\dimPatchesFigs}{0.149\linewidth}%
\newcommand{\trimPatches}{0.09\linewidth}%
\newcommand{\drawRectPatches}[1]{\begin{tikzpicture}
        \node[anchor=south west,inner sep=0] (image) at (0,0) {\includegraphics[width=\dimPatchesFigs, height=\dimPatchesFigs]{#1}};
        \begin{scope}[x={(image.south east)},y={(image.north west)}]
            \draw[red,ultra thick,rounded corners] (0.15,0.15) rectangle (0.85,0.85);
        \end{scope}
    \end{tikzpicture}
}%
\newcommand{\plotPatchesFigsWithRectCaptions}[1]{\captionsetup[subfloat]{position=bottom}
    \setcounter{subfigure}{0} 
    \subfloat[GT]{\drawRectPatches{figs/results/patches/#1/full_4/label.pdf}}
    \hfill
    \subfloat[Sample $\times 2$]{\includegraphics[trim={{\trimPatches} {\trimPatches} {\trimPatches} {\trimPatches}}, clip=True, width=\dimPatchesFigs, height=\dimPatchesFigs]{figs/results/patches/#1/full_2/sample.pdf}}
    \hfill
    \subfloat[SR $\times 2$]{\includegraphics[trim={{\trimPatches} {\trimPatches} {\trimPatches} {\trimPatches}}, clip=True, width=\dimPatchesFigs, height=\dimPatchesFigs]{figs/results/patches/#1/full_2/SingleSR.pdf}}
    \hfill
    \subfloat[Sample $\times 4$]{\includegraphics[trim={{\trimPatches} {\trimPatches} {\trimPatches} {\trimPatches}}, clip=True, width=\dimPatchesFigs, height=\dimPatchesFigs]{figs/results/patches/#1/full_4/sample.pdf}}
    \hfill
    \subfloat[SR $\times 4$]{\includegraphics[trim={{\trimPatches} {\trimPatches} {\trimPatches} {\trimPatches}}, clip=True, width=\dimPatchesFigs, height=\dimPatchesFigs]{figs/results/patches/#1/full_4/SingleSR.pdf}}
    \hfill
    \subfloat[GT (patch)]{\includegraphics[trim={{\trimPatches} {\trimPatches} {\trimPatches} {\trimPatches}}, clip=True, width=\dimPatchesFigs, height=\dimPatchesFigs]{figs/results/patches/#1/full_4/label.pdf}}
}%
\newcommand{\plotPatchesFigsWithRect}[1]{\subfloat{\drawRectPatches{figs/results/patches/#1/full_4/label.pdf}}
    \hfill
    \subfloat{\includegraphics[trim={{\trimPatches} {\trimPatches} {\trimPatches} {\trimPatches}}, clip=True, width=\dimPatchesFigs, height=\dimPatchesFigs]{figs/results/patches/#1/full_2/sample.pdf}}
    \hfill
    \subfloat{\includegraphics[trim={{\trimPatches} {\trimPatches} {\trimPatches} {\trimPatches}}, clip=True, width=\dimPatchesFigs, height=\dimPatchesFigs]{figs/results/patches/#1/full_2/SingleSR.pdf}}
    \hfill
    \subfloat{\includegraphics[trim={{\trimPatches} {\trimPatches} {\trimPatches} {\trimPatches}}, clip=True, width=\dimPatchesFigs, height=\dimPatchesFigs]{figs/results/patches/#1/full_4/sample.pdf}}
    \hfill
    \subfloat{\includegraphics[trim={{\trimPatches} {\trimPatches} {\trimPatches} {\trimPatches}}, clip=True, width=\dimPatchesFigs, height=\dimPatchesFigs]{figs/results/patches/#1/full_4/SingleSR.pdf}}
    \hfill
    \subfloat{\includegraphics[trim={{\trimPatches} {\trimPatches} {\trimPatches} {\trimPatches}}, clip=True, width=\dimPatchesFigs, height=\dimPatchesFigs]{figs/results/patches/#1/full_4/label.pdf}}
}%
\begin{figure*}
    \centering
    \plotPatchesFigsWithRect{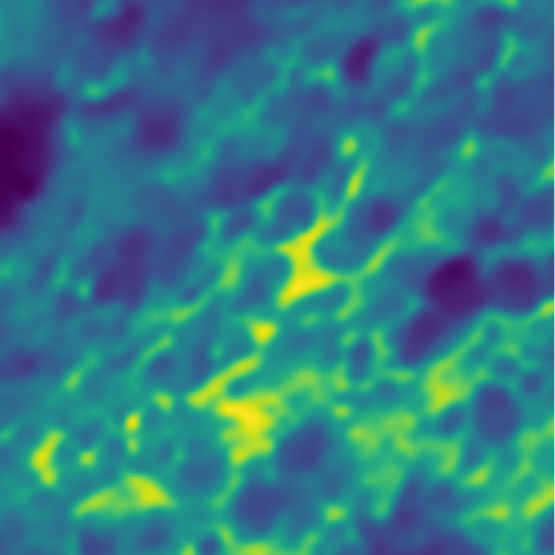}\\
    \plotPatchesFigsWithRect{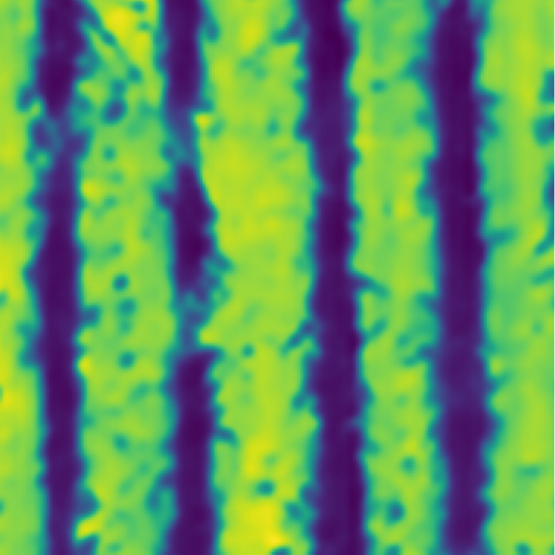}\\
    \plotPatchesFigsWithRect{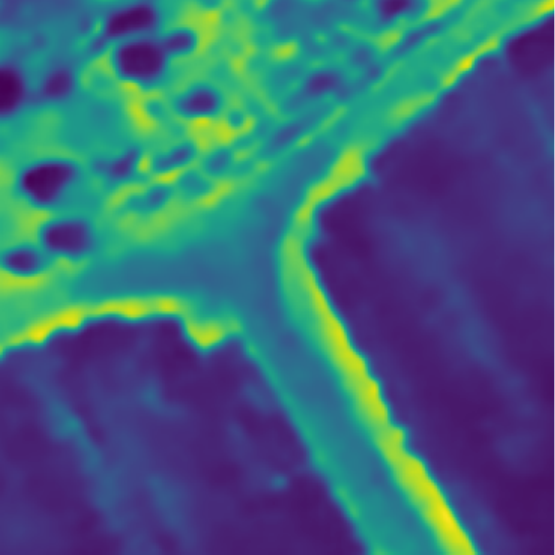}\\
    \plotPatchesFigsWithRect{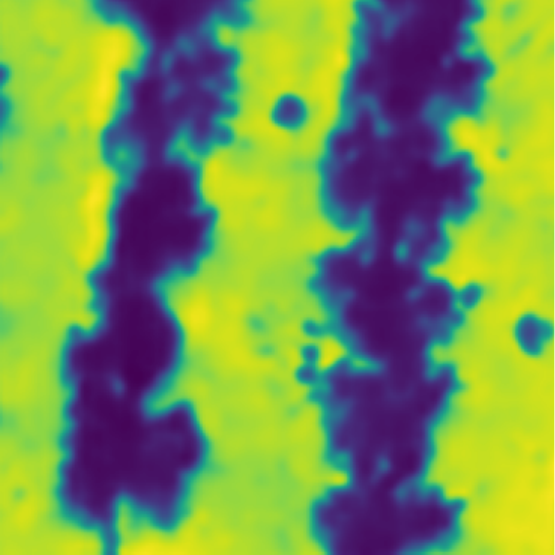}\\
    \plotPatchesFigsWithRectCaptions{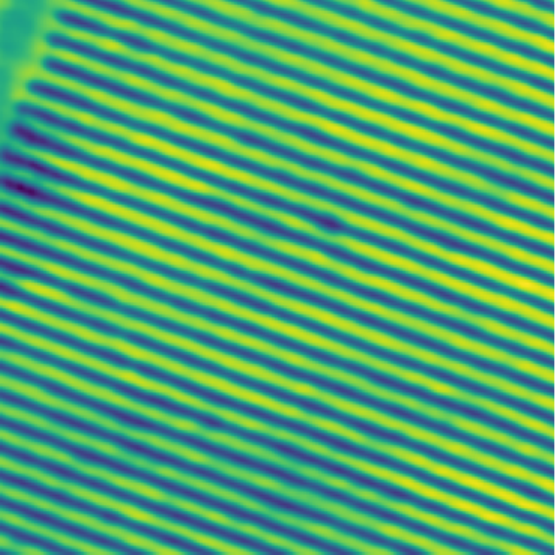}
    \caption{\captionResultsSimPatches}
    \label{fig:results:sim:patches:1}
\end{figure*}%
\begin{figure*}
    \centering
    \plotPatchesFigsWithRect{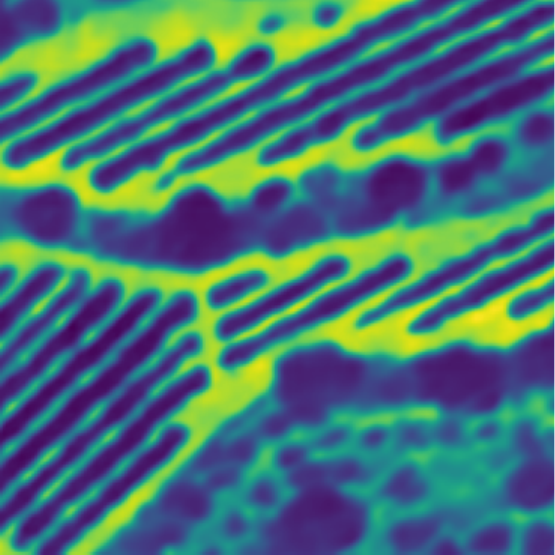}\\
    \plotPatchesFigsWithRect{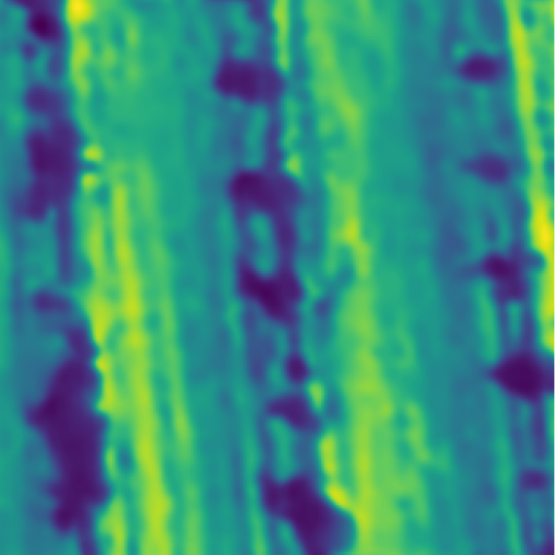}\\
    \plotPatchesFigsWithRect{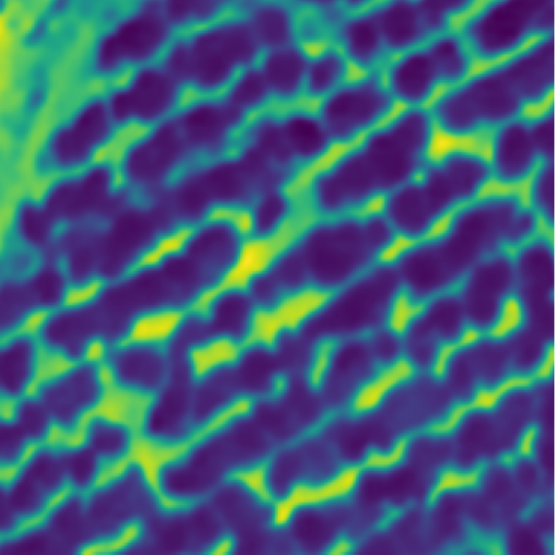}\\
    \plotPatchesFigsWithRect{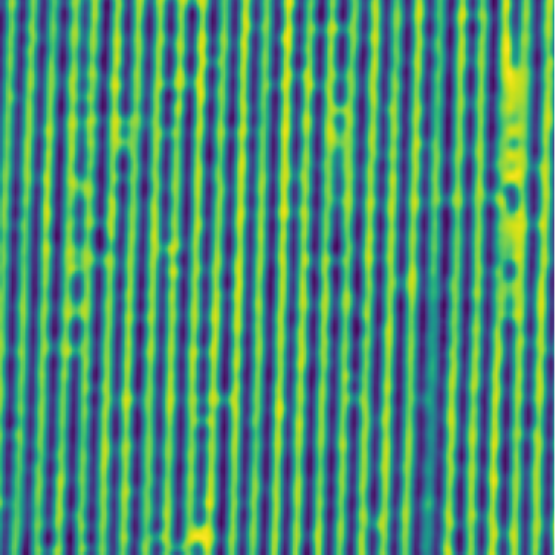}\\
    \plotPatchesFigsWithRectCaptions{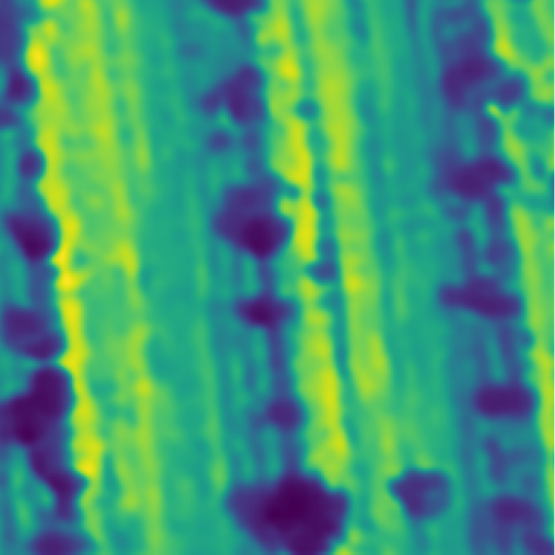}
    \caption{\captionResultsSimPatches}
    \label{fig:results:sim:patches:2}
\end{figure*}%
\subsubsection{Qualitative results} \cref{fig:results:sim:patches:1,fig:results:sim:patches:2} show the results of the end-to-end pipeline for both scale factors ($\times2$ and $\times4$).

To illustrate the method qualitatively, the authors reduced the resolution of each ground truth image by factors of X2 and X4 and then restored them to their original size, enabling a comparison with the ground truth. The resolution reduction simulated aerial imaging from twice and four times the original height, respectively. This increased the field of view of each frame by X4 and X16, respectively, which has agricultural benefits such as reducing the flight time over a given field or increasing the land coverage with the same effort, implying cost savings.

For each figure, the left-most column (a) is the GT\@. The red rectangle in the GT is the area enlarged in the following subfigures.
The second and third columns (b and c) show the sample downscaled by $\times2$ and the SR results for $\times2$ upscaling, respectively; the fourth column (d) shows the sample downscaled by $\times4$ and the fifth column (e) shows the SR results for $\times4$ upscaling; the last column (f) shows the GT patch.
All results were obtained with 11 frames for the multiframe NUC (\cref{sec:methods:nuc:multi}).

Although the $\times4$ sample (e) contained very few details due to the LR, the SR was still able to achieve good results.
Some of the samples even suffered from aliasing, and the SR still managed to provide a good estimate (e.g., \cref{fig:results:sim:patches:1} third to last and last columns (d and f).).

\subsubsection{Line-set comparison}
To demonstrate the accuracy of the temperature estimation, \cref{fig:results:sim:lines} display the results of the end-to-end pipeline and the GT as the middle horizontal line of the frame.
The results are shown for both scale factors ($\times2$ and $\times4$).
On each graph the black line is the GT, the blue line is the end-to-end pipeline with the single-frame NUC, and the red line is the end-to-end pipeline with the multiframe NUC.
\newcommand{\linePlotLegendFontSize}{\tiny}%
\newcommand{\locLinePlotLegend}{1.65}%
\newcommand{\heightLinePlot}{0.36\linewidth}%
\newcommand{\sizesSubfloatsLinePlots}{0.41\linewidth}%
\newcommand{\sizesFontLinePlots}{\footnotesize}%
\newcommand{\plotLineSetsNoLegend}[1]{\addplot[color=blue] table [x=x, y=single, col sep=comma] {#1};
    \addplot[color=red] table [x=x, y=multi, col sep=comma] {#1};
    \addplot[color=black] table [x=x, y=gt, col sep=comma] {#1};
}%
\newcommand{\plotLineSets}[1]{\addplot[color=blue] table [x=x, y=single, col sep=comma] {#1};
    \addlegendentry{SR + Single-frame NUC}
    \addplot[color=red] table [x=x, y=multi, col sep=comma] {#1};
    \addlegendentry{SR + Multiframe NUC}
    \addplot[color=black] table [x=x, y=gt, col sep=comma] {#1};
    \addlegendentry{GT}
}%
\newcommand{\plotLineSetsAxisXY}[3]{\begin{tikzpicture}
        \begin{axis}[
                width=#1,
                height=\heightLinePlot,
                font=#2,
                xlabel={Index of pixel},
                ylabel={Temperature [$^\circ \text{C}$]},
                legend style={at={(\locLinePlotLegend,0.98)}, anchor=north east, font=\linePlotLegendFontSize},
                grid
            ]
            \plotLineSets{#3}
        \end{axis}
    \end{tikzpicture}
}%
\newcommand{\plotLineSetsAxisX}[3]{\begin{tikzpicture}
        \begin{axis}[
                width=#1,
                height=\heightLinePlot,
                font=#2,
                xlabel={Index of pixel},
                legend style={at={(0.98,0.98)}, anchor=north east, font=\linePlotLegendFontSize},
                grid
            ]
            \plotLineSetsNoLegend{#3}
        \end{axis}
    \end{tikzpicture}
}%
\newcommand{\plotLineSetsAxisY}[3]{\begin{tikzpicture}
        \begin{axis}[
                width=#1,
                height=\heightLinePlot,
                font=#2,
                ylabel={Temperature [$^\circ \text{C}$]},
                legend style={at={(\locLinePlotLegend,0.98)}, anchor=north east, font=\linePlotLegendFontSize},
                grid
            ]
            \plotLineSets{#3}
        \end{axis}
    \end{tikzpicture}
}%
\newcommand{\plotLineSetsNoAxis}[3]{\begin{tikzpicture}
        \begin{axis}[
                width=#1,
                height=\heightLinePlot,
                font=#2,
                legend style={at={(0.98,0.98)}, anchor=north east, font=\linePlotLegendFontSize},
                grid
            ]
            \plotLineSetsNoLegend{#3}
        \end{axis}
    \end{tikzpicture}
}%
\begin{figure*}
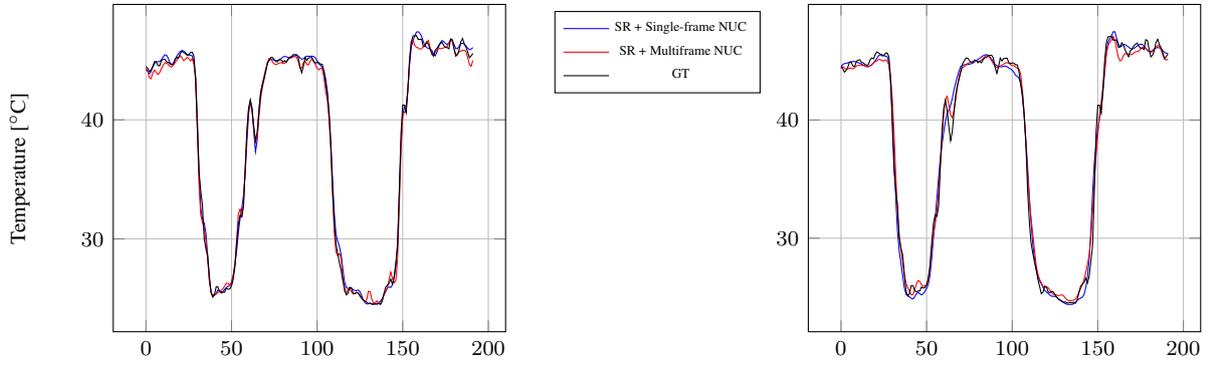
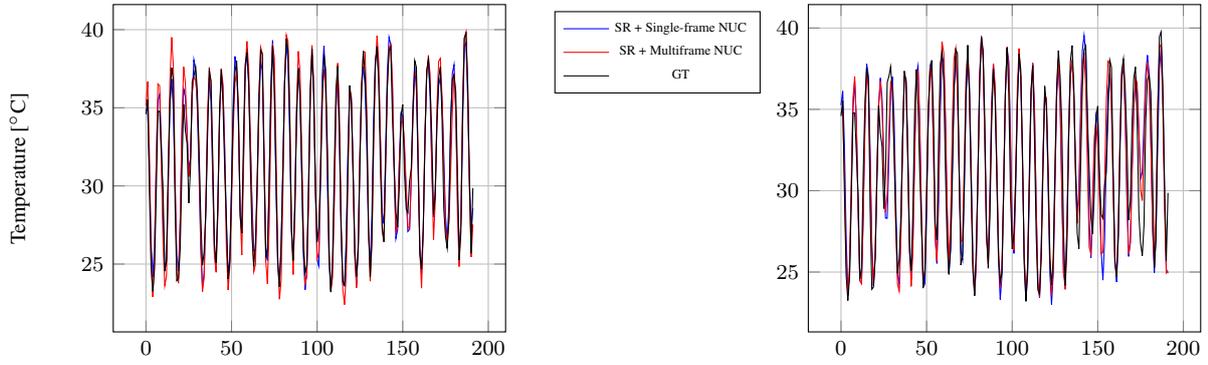
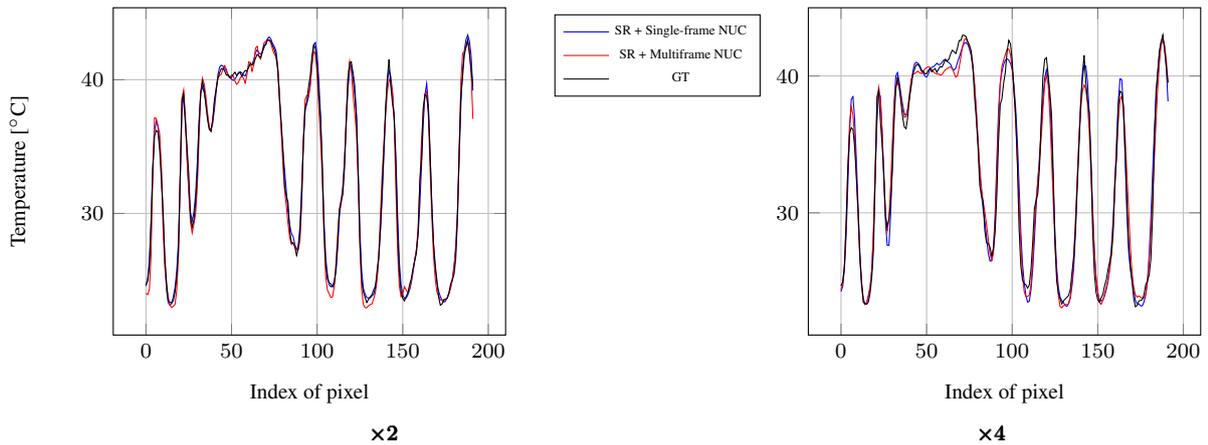
\label{fig:results:sim:lines}
    \centering
    \subfloat[Tzuba, 2018-Aug-05]{\begin{minipage}{\linewidth}
            \captionsetup{type=figure}
            \subfloat{\plotLineSetsAxisY{\sizesSubfloatsLinePlots}{\sizesFontLinePlots}{figs/results/patches/NEW0063360/lines_2.csv}}
            \subfloat{\plotLineSetsNoAxis{\sizesSubfloatsLinePlots}{\sizesFontLinePlots}{figs/results/patches/NEW0063360/lines_4.csv}}%
        \end{minipage}\setcounter{subfigure}{1}}
    \\
    \subfloat[Tzora, 2018-Aug-05]{\begin{minipage}{\linewidth}
            \captionsetup{type=figure}
            \subfloat{\plotLineSetsAxisY{\sizesSubfloatsLinePlots}{\sizesFontLinePlots}{figs/results/patches/Tzora735/lines_2.csv}}
            \subfloat{\plotLineSetsNoAxis{\sizesSubfloatsLinePlots}{\sizesFontLinePlots}{figs/results/patches/Tzora735/lines_4.csv}}%
        \end{minipage}\setcounter{subfigure}{2}}
    \\
    \subfloat[Barkan, 2019-Aug-12]{\begin{minipage}{\linewidth}
            \captionsetup{type=figure}
            \setcounter{subfigure}{0}
            \captionsetup[subfloat]{labelformat=empty}
            \subfloat[$\pmb{\times2}$]{\plotLineSetsAxisXY{\sizesSubfloatsLinePlots}{\sizesFontLinePlots}{figs/results/patches/barkan_190812_6905/lines_2.csv}}
            \subfloat[$\pmb{\times4}$]{\plotLineSetsAxisX{\sizesSubfloatsLinePlots}{\sizesFontLinePlots}{figs/results/patches/barkan_190812_6905/lines_4.csv}}%
        \end{minipage}\setcounter{subfigure}{3}}
    \caption{Values of the horizontal line in the middle of the frame are for the ground-truth (GT) $\frameGT$, the proposed end-to-end pipeline when using single-frame nonuniformity correction (NUC) and the proposed end-to-end pipeline when using multiframe NUC\@. Left column: $\times2$, right column $\times4$ temperature estimations, respectively.}
\end{figure*}%

The figures exemplify the high quality of the SR temperature estimation after the end-to-end pipeline, by visually demonstrating that the results are very similar to the GT.

\subsubsection{Runtime comparison}\label{sec:results:sim:runtime}
\cref{table:results:sim:runtime} shows the frames processed per second (FPS) by the different methods.
The runtime was measured over the entire validation dataset, and averaged by the number of frames.
The CPU used was i9-13900k, and the GPU was 4070Ti.

All methods were able to process frames in less than a second on a consumer-grade CPU, making the proposed method suitable for real-time video applications.
For the GPU, the times were significantly faster, on-par with the FPS of the camera itself.
Notice that the multiframe NUC was slower than the single-frame NUC, as expected, due to the additional processing required to process multiple frames simultaneously.

The $\times2$ upscale was slower than the $\times4$ upscale for both single- and multiframe NUC\@.
This means that the SR module's impact on runtime is negligible compared to that of the NUC module.
That is, when applying the NUC module on frames with smaller spatial dimensions, the runtime is affected considerably more than by the extra calculations that the SR module requires for upscaling to $\times4$ rather than $\times2$.
\begin{table}[H]
    \centering
    \caption{Runtime of the end-to-end pipeline for $\times2$ and $\times4$ SR with single and multiframe nonuniformity correction (NUC)\@.
        Values are the number of frames processed per second.
        The CPU used was i9-13900k, and the GPU was 4070Ti.}
    \begin{tabular}{|c||c|c||c|c|}
        \hline
        Scale factor & \multicolumn{2}{|c||}{2} & \multicolumn{2}{|c|}{4}                  \\
        \hline
        Type         & Single                   & Multi                   & Single & Multi \\
        \hline
        \hline
        CPU          & 6.3                      & 3.7                     & 18.1   & 12.3  \\ \hline
        GPU          & 37.1                     & 38.9                    & 65.2   & 49.0  \\ \hline
    \end{tabular}
    \label{table:results:sim:runtime}
\end{table}%
\subsection{Real data}\label{sec:results:real}
\newcommand{\realdataFigHeight}{0.23\linewidth}%
\newcommand{\realdataFigWidth}{0.23\linewidth}%
\newenvironment{realDataFigures}[2]{
    \CatchFileDef{\singleTwo}{figs/results/realdata/#1/#2/single_2.txt}{}
    \CatchFileDef{\singleFour}{figs/results/realdata/#1/#2/single_4.txt}{}
    \CatchFileDef{\multiTwo}{figs/results/realdata/#1/#2/multi_2.txt}{}
    \CatchFileDef{\multiFour}{figs/results/realdata/#1/#2/multi_4.txt}{}
        \centering
        \subfloat{\begin{tikzpicture}
            \node[anchor=south west,inner sep=0] (image) at (0,0) {\includegraphics[height=\realdataFigHeight]{figs/results/realdata/#1/#2/single_2.pdf}};
            \begin{scope}[x={(image.south east)},y={(image.north west)}]
                \node[fill=black,text=white] at (0.15,0.90) {\singleTwo};
            \end{scope}
        \end{tikzpicture}}
        \hfill
        \subfloat{\begin{tikzpicture}
            \node[anchor=south west,inner sep=0] (image) at (0,0) {\includegraphics[height=\realdataFigHeight]{figs/results/realdata/#1/#2/single_4.pdf}};
            \begin{scope}[x={(image.south east)},y={(image.north west)}]
                \node[fill=black,text=white] at (0.15,0.90) {\singleFour};
            \end{scope}
        \end{tikzpicture}}
        \hfill
        \subfloat{\begin{tikzpicture}
            \node[anchor=south west,inner sep=0] (image) at (0,0) {\includegraphics[height=\realdataFigHeight]{figs/results/realdata/#1/#2/multi_2.pdf}};
            \begin{scope}[x={(image.south east)},y={(image.north west)}]
                \node[fill=black,text=white] at (0.15,0.90) {\multiTwo};
            \end{scope}
        \end{tikzpicture}}
        \hfill
        \subfloat{\begin{tikzpicture}
            \node[anchor=south west,inner sep=0] (image) at (0,0) {\includegraphics[height=\realdataFigHeight]{figs/results/realdata/#1/#2/multi_4.pdf}};
            \begin{scope}[x={(image.south east)},y={(image.north west)}]
                \node[fill=black,text=white] at (0.15,0.90) {\multiFour};
            \end{scope}
        \end{tikzpicture}}
        \hfill
        \subfloat{\includegraphics[height=\realdataFigHeight]{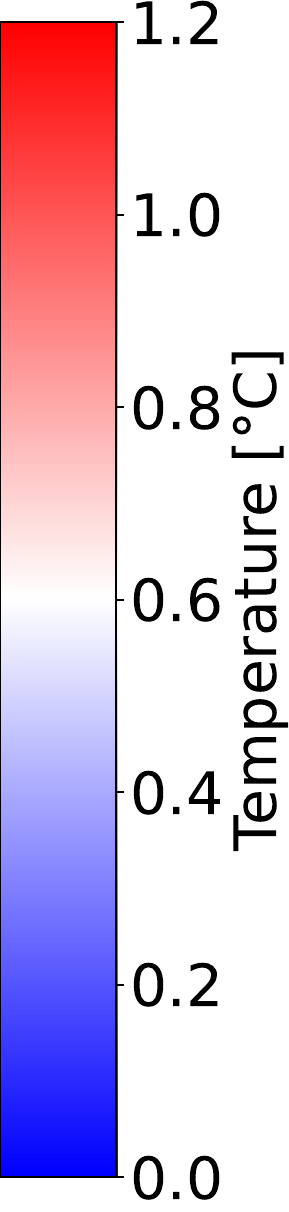}}
        }{}%
\newenvironment{realDataFiguresWithCaptions}[2]{
    \setcounter{subfigure}{0}
    \CatchFileDef{\singleTwo}{figs/results/realdata/#1/#2/single_2.txt}{}
    \CatchFileDef{\singleFour}{figs/results/realdata/#1/#2/single_4.txt}{}
    \CatchFileDef{\multiTwo}{figs/results/realdata/#1/#2/multi_2.txt}{}
    \CatchFileDef{\multiFour}{figs/results/realdata/#1/#2/multi_4.txt}{}
        \centering
        \subfloat[Single $\times2$]{\begin{tikzpicture}
            \node[anchor=south west,inner sep=0] (image) at (0,0) {\includegraphics[width=\realdataFigWidth, height=\realdataFigHeight]{figs/results/realdata/#1/#2/single_2.pdf}};
            \begin{scope}[x={(image.south east)},y={(image.north west)}]
                \node[fill=black,text=white] at (0.15,0.90) {\singleTwo};
            \end{scope}
        \end{tikzpicture}}
        \hfill
        \subfloat[Single $\times4$]{\begin{tikzpicture}
            \node[anchor=south west,inner sep=0] (image) at (0,0) {\includegraphics[width=\realdataFigWidth, height=\realdataFigHeight]{figs/results/realdata/#1/#2/single_4.pdf}};
            \begin{scope}[x={(image.south east)},y={(image.north west)}]
                \node[fill=black,text=white] at (0.15,0.90) {\singleFour};
            \end{scope}
        \end{tikzpicture}}
        \hfill
        \subfloat[Multi $\times2$]{\begin{tikzpicture}
            \node[anchor=south west,inner sep=0] (image) at (0,0) {\includegraphics[width=\realdataFigWidth, height=\realdataFigHeight]{figs/results/realdata/#1/#2/multi_2.pdf}};
            \begin{scope}[x={(image.south east)},y={(image.north west)}]
                \node[fill=black,text=white] at (0.15,0.90) {\multiTwo};
            \end{scope}
        \end{tikzpicture}}
        \hfill
        \subfloat[Multi $\times4$]{\begin{tikzpicture}
            \node[anchor=south west,inner sep=0] (image) at (0,0) {\includegraphics[width=\realdataFigWidth, height=\realdataFigHeight]{figs/results/realdata/#1/#2/multi_4.pdf}};
            \begin{scope}[x={(image.south east)},y={(image.north west)}]
                \node[fill=black,text=white] at (0.15,0.90) {\multiFour};
            \end{scope}
        \end{tikzpicture}}
        \hfill
        \subfloat{\includegraphics[height=\realdataFigHeight]{figs/results/realdata/colorbar.pdf}}
        }{}%
\begin{figure*}
    \begin{realDataFiguresWithCaptions}{main}{31}
    \end{realDataFiguresWithCaptions}\\
    \begin{realDataFiguresWithCaptions}{main}{55}
    \end{realDataFiguresWithCaptions}\\
    \begin{realDataFiguresWithCaptions}{main}{77}
    \end{realDataFiguresWithCaptions}\\
    \begin{realDataFiguresWithCaptions}{main}{500}
    \end{realDataFiguresWithCaptions}
    \caption{Results of the proposed end-to-end pipeline on real data.
    The ground-truth (GT) temperature map is in gray, and superimposed on it is the error between the estimated temperature map and the GT temperature map.
    The number in white at the top-left of each panel is the mean absolute error (MAE) between the estimated temperature map and the GT temperature map.}
    \label{fig:results:real:1}
\end{figure*}%
\begin{figure*}
    \begin{realDataFiguresWithCaptions}{main}{60}
    \end{realDataFiguresWithCaptions}\\
    \begin{realDataFiguresWithCaptions}{main}{22}
    \end{realDataFiguresWithCaptions}\\
    \begin{realDataFiguresWithCaptions}{main}{57}
    \end{realDataFiguresWithCaptions}\\
    \begin{realDataFiguresWithCaptions}{main}{91}
    \end{realDataFiguresWithCaptions}
    \caption{Results of the proposed end-to-end pipeline on real data.
    The ground-truth (GT) temperature map is in gray, and superimposed on it is the error between the estimated temperature map and the GT temperature map.
    The number in white at the top-left of each panel is the mean absolute error (MAE) between the estimated temperature map and the GT temperature map.}
    \label{fig:results:real:2}
\end{figure*}%
To validate the proposed pipeline on real data, \taucamera and \scientificCamera cameras were mounted together on a UAV\@.
To simulate scaling between the two cameras, the focal length of the \scientificCamera was set to $\times 2$ that of the \taucamera's focal length.
\cref{appendix:fig:uav} shows the UAV with the two cameras mounted on it.
The frames were collected over Beit Dagan in Israel (coordinates $32.0026, 34.8154$).
The entire dataset is available at the end of the manuscript, in the data availability section.
After collecting the frames, the raw gray-level low-resolution \taucamera frames were processed by the proposed end-to-end pipeline (\cref{fig:methods:e2e}) to create estimated HR temperature maps.
The estimated temperature maps were registered by hand to the GT HR \scientificCamera frames, resulting in a dataset of 46 registered frames.
Due to technical difficulties, only a single flight altitude was collected. Thus, the \scientificCamera GT is $\times2$ of the $\taucamera$'s resolution. To create a $\times4$ dataset, the GT frames were downscaled by a factor of 2 using a bicubic interpolation.

Some of the results of applying the end-to-end pipeline on the real data are shown in \cref{fig:results:real:1,fig:results:real:2}.
The figures display the accurate GT temperature map in gray, and superimposed on it is the error between the estimated temperature map and the GT temperature map in color. The error map is smaller due to the differences in the field of view between the \taucamera and the \scientificCamera.
The number in white at the top left of each panel is the MAE between the estimated temperature map and the GT temperature map in degrees Celsius.
The average MAEs across the 46 frames were:
\begin{itemize}
    \item $\times2$ single-frame NUC: $0.92^\circ$C
    \item $\times2$ multiframe NUC: $0.81^\circ$C
    \item $\times4$ single-frame NUC: $0.86^\circ$C
    \item $\times4$ multiframe NUC: $0.81^\circ$C
\end{itemize}

The results demonstrate that the proposed pipeline can produce an accurate estimation of HR temperature maps from LR raw gray-level frames.
In the results displayed in \cref{fig:results:real:1,fig:results:real:2} the estimated temperature maps are visually similar to the GT temperature maps (blue indicates lower error).
Moreover, the average MAE is on the order of sub-degrees.
Notice that the error is also affected by the quality of the registration, making the actual accuracy of the proposed pipeline is even better.
\subsection{Crop water stress index (CWSI)}\label{sec:results:cwsi}
\newcommand{\dimCwsiPatches}{0.205\linewidth}%
\newcommand{\trimCwsiPatches}{0.135\linewidth}%
\newcommand{\drawRectCwsi}[1]{\begin{tikzpicture}
        \node[anchor=south west,inner sep=0] (image) at (0,0) {\includegraphics[width=\dimCwsiPatches, height=\dimCwsiPatches]{#1}};
        \begin{scope}[x={(image.south east)},y={(image.north west)}]
            \draw[red,ultra thick,rounded corners] (0.2, 0.2) rectangle (0.8,0.8);
        \end{scope}
    \end{tikzpicture}
}%
\newcommand{\plotCwsiFigs}[2]{\subfloat{\drawRectCwsi{figs/results/cwsi/#1/#2/single/label.pdf}}
    \hfill
    \subfloat{\includegraphics[trim={{\trimCwsiPatches} {\trimCwsiPatches} {\trimCwsiPatches} {\trimCwsiPatches}}, clip=True, width=\dimCwsiPatches, height=\dimCwsiPatches]{figs/results/cwsi/#1/#2/single/sample.pdf}}
    \hfill
    \subfloat{\includegraphics[trim={{\trimCwsiPatches} {\trimCwsiPatches} {\trimCwsiPatches} {\trimCwsiPatches}}, clip=True, width=\dimCwsiPatches, height=\dimCwsiPatches]{figs/results/cwsi/#1/#2/single/sr.pdf}}
    \hfill
    \subfloat{\includegraphics[trim={{\trimCwsiPatches} {\trimCwsiPatches} {\trimCwsiPatches} {\trimCwsiPatches}}, clip=True, width=\dimCwsiPatches, height=\dimCwsiPatches]{figs/results/cwsi/#1/#2/single/error_cwsi.pdf}}
}%
\newcommand{\plotCwsiFigsCaption}[2]{\setcounter{subfigure}{0} 
    \subfloat[GT]{\drawRectCwsi{figs/results/cwsi/#1/#2/single/label.pdf}}
    \hfill
    \subfloat[Sample $\times{#2}$]{\includegraphics[trim={{\trimCwsiPatches} {\trimCwsiPatches} {\trimCwsiPatches} {\trimCwsiPatches}}, clip=True, width=\dimCwsiPatches, height=\dimCwsiPatches]{figs/results/cwsi/#1/#2/single/sample.pdf}}
    \hfill
    \subfloat[Ours $\times{#2}$]{\includegraphics[trim={{\trimCwsiPatches} {\trimCwsiPatches} {\trimCwsiPatches} {\trimCwsiPatches}}, clip=True, width=\dimCwsiPatches, height=\dimCwsiPatches]{figs/results/cwsi/#1/#2/single/sr.pdf}}
    \hfill
    \subfloat[CWSI $\times{#2}$]{\includegraphics[trim={{\trimCwsiPatches} {\trimCwsiPatches} {\trimCwsiPatches} {\trimCwsiPatches}}, clip=True, width=\dimCwsiPatches, height=\dimCwsiPatches]{figs/results/cwsi/#1/#2/single/error_cwsi.pdf}}
}%
\newcommand{\cwsiFigCaptions}[1]{Crop water stress index (CWSI) results for the ground-truth (GT) and for the proposed end-to-end pipeline for temperature estimation and super resolution (SR).
    (a) GT; (b) #1 downsampled sample (enlarged for preview); (c) SR results; (d) CWSI\@.
    The CWSI results obtained from the proposed method are in green, and those from the GT are in red.
    The red is almost unnoticeable because of the small error between the GT and estimated
}%
\begin{figure*}
    \centering
    \plotCwsiFigs{Almonds_210722_1}{2}\\
    \plotCwsiFigs{MevoBytar_200805}{2}\\
    \plotCwsiFigs{Hohova_220629}{2}\\
    \plotCwsiFigs{MevoBytar_200722}{2}\\
    \plotCwsiFigsCaption{Avocado_230809}{2}
    \caption{\cwsiFigCaptions{$\times2$}}
    \label{fig:results:cwsi:scale2}
\end{figure*}%
\begin{figure*}
    \centering
    \plotCwsiFigs{Almonds_210722_1}{4}\\
    \plotCwsiFigs{MevoBytar_200805}{4}\\
    \plotCwsiFigs{Hohova_220629}{4}\\
    \plotCwsiFigs{MevoBytar_200722}{4}\\
    \plotCwsiFigsCaption{Avocado_230809}{4}
    \caption{\cwsiFigCaptions{$\times4$}}
    \label{fig:results:cwsi:scale4}
\end{figure*}%
\begin{table*}
    \centering
    \caption{Percentage of error between the crop water stress index (CWSI) from the ground-truth (GT) thermal frame and the CWSI results from the proposed method.
        The results for single- and multiframe images are shown for both $\times2$ and $\times4$ super resolution.
        All images were taken in Israel with a \scientificCamera camera mounted on a UAV\@.
        The multiframe results were obtained from 7 consecutive frames.
        The dimensions of all frames were $1024\times1024$ pixels.
        "X" - the algorithm was unable to provide an estimation due to some unknown issue on that specific GT frame.}
    \begin{tabular}{|c|c|c|c|c||c|c||c|c|}
        \hline
        \multirow{2}{*}{Location} & \multirow{2}{*}{Crop} & \multirow{2}{*}{Date} & \multirow{2}{*}{Time [AM]} & Ambient                 & \multicolumn{2}{|c|}{Single-frame} & \multicolumn{2 }{|c|}{Multiframe}                         \\
        \cline{6-9}
                                  &                       &                       &                            & temperature [$^\circ$C] & $\times2$                          & $\times4$                          & $\times2$ & $\times4$ \\
        \hline
        \hline
        Sede Nehamia              & Almonds (2)           & 2021-Jul-22           & 10                         & 34                      & 2\%                                & 3\%                                & 8\%       & 4\%       \\\hline

        Sede Nehamia              & Almonds (1)           & 2021-Jul-22           & 9                          & 33                      & 3\%                                & 3\%                                & 8\%       & 2\%       \\\hline

        Gevim                     & Avocado               & 2023-Aug-09           & 9                          & 32                      & 1\%                                & 1\%                                & 4\%       & 7\%       \\\hline

        Kfar Habad                & Citrus                & 2023-Jun-07           & 8                          & 28.6                    & 1\%                                & 3\%                                & 9\%       & 2\%       \\\hline

        Sede David                & Jojoba                & 2022-Jun-29           & 9                          & 30.5                    & 1\%                                & 1\%                                & 2\%       & 4\%       \\\hline

        Mevo Bytar                & Grapes (2)            & 2020-Jul-22           & 9                          & 30                      & 1\%                                & 1\%                                & 9\%       & 2\%       \\\hline

        Mevo Bytar                & Grapes (1)            & 2020-Aug-05           & 10                         & 25.2                    & 1\%                                & 1\%                                & X         & 3\%       \\\hline
    \end{tabular}
    \label{table:results:cwsi}
\end{table*}%
\cref{table:results:cwsi} shows the difference between the CWSI values calculated on the GT and on the estimation for the different methods (single-frame and multiframe NUC for both $\times2$ and $\times4$ SR). The results are displayed as percent error in CWSI between the GT and the estimation. The GT frames were $1024\times1024$ pixels, and the ambient temperature was acquired by a portable meteorological station.

The results in the table indicate that the proposed end-to-end pipeline does not introduce any significant error in the CWSI estimation, which is very close to the GT\@. This finding is consistent with the real data results in \cref{sec:results:real} and simulated results in \cref{sec:results:sim}, suggesting that the proposed method can be used to estimate the CWSI in real scenarios.

\cref{fig:results:cwsi:scale2,fig:results:cwsi:scale4} display some results of the CWSI estimation for both scale factors ($\times2$ and $\times4$).
The high resolution GT temperature map is shown in the left-most column (a), the second column (b) is the input sample downscaled by either $\times 2$ for \cref{fig:results:cwsi:scale2} or $\times 4$ for \cref{fig:results:cwsi:scale4}, the third column (c) is the output of the proposed end-to-end pipeline, and the right-most column (d) is the CWSI estimation in green, superimposed on the estimated temperature map.
For both figures, the CWSI for the GT temperature map is superimposed in red. Notice that for \cref{fig:results:cwsi:scale2}, the red is almost unnoticeable, meaning that the GT and estimated CWSI are very close. In \cref{fig:results:cwsi:scale4}, the first row does contain some red pixels, but the estimated CWSI is still very close to the GT CWSI - mean errors of $1.42\%$ for $\times2$ and $1.86\%$ for $\times4$ scale factors.
The results are consistent with the quantitative results in \cref{table:results:cwsi}.
\section{Discussion}\label{sec:discussion}
An end-to-end pipeline for temperature estimation and SR of frames captured by a low-cost uncooled IR camera is presented. The pipeline consists of two main components: a deep-learning-based temperature estimation module that learned to map the raw gray-level IR images to the corresponding temperature maps while also correcting for nonuniformity, and a deep-learning-based SR module which uses a deep-learning network to enhance the spatial resolution of the IR images by $\times2$ and $\times4$. The performance of the pipeline was evaluated on both simulated and real-world datasets. Accurate temperature estimation was indeed generated by the proposed end-to-end pipeline.

To mimic a low-cost IR camera, the camera features were disabled and only a single-point correction was applied to the raw frames - similar to common practices in field work. This procedure resulted in basic unprocessed gray level frames being acquired by the camera.
Using these basic frames, a simulator of low-cost IR cameras was developed to generate simulated IR frames and temperature maps of various scenes (\cref{sec:methods:data}).
The simulated dataset was used to train and validate the results of the proposed pipeline.
For the simulated dataset (\cref{sec:results:sim}) the average MAE for the single-frame NUC was $0.54^\circ\text{C}$ for $\times2$ SR and $0.93^\circ\text{C}$ for $\times4$ SR, and the average MAE for the multiframe NUC was $0.57^\circ\text{C}$ for $\times2$ SR and $0.84^\circ\text{C}$ for $\times4$ SR.
The runtime for the end-to-end pipeline on a CPU was less than $1_{\text{sec}}$ per frame, and on GPU at real-time video rate (\cref{sec:results:sim:runtime}).

A real-world dataset was collected using a UAV\@.
Low-cost IR frames and accurate temperature maps of the same scenes were captured (\cref{sec:results:real}).
The proposed end-to-end pipeline was applied to the IR frames to produce estimated temperature maps, which were registered and compared to the GT temperature maps (\cref{fig:results:real:1,fig:results:real:2}).
The results were accurate at the sub-degree level, and the visual results showed that accurate temperature maps from low-cost IR frames can be produced by the proposed end-to-end pipeline, and that it can compete with high-end thermal cameras in terms of quality and accuracy of the temperature estimation.

The CWSI values estimated from the output of the proposed pipeline were accurate relative to the GT CWSI (\cref{sec:results:cwsi}).
The average error between the GT and the estimated CWSI for the single-frame NUC was $1.42\%$ for $\times2$ and $1.86\%$ for $\times4$, and the average error for the multiframe NUC was $6.67\%$ for $\times2$ and $3.43\%$ for $\times4$.
\section{Conclusion}\label{sec:conclusion}
In conclusion, we propose a pipeline for temperature estimation and SR of frames captured by a low-cost uncooled IR camera.

The pipeline has several advantages over existing methods for temperature estimation and SR of IR images.
First, it is end-to-end, meaning that no pre- or post-processing of the IR images is required. 
Second, it is data-driven, meaning that it can be adapted to different IR cameras, scenes and temperature ranges, without requiring any manual tuning of the parameters.
Third, it is scalable - it can handle different input and output resolutions, depending on the application and the hardware specifications. 
Fourth, it can be run in real-time at video rates on a GPU, or at a high throughput of more than 1 frames per second on a CPU, enabling online processing, or fast offline processing.

This pipeline was validated on both real-world and simulated data, and produced accurate temperature estimations with higher spatial resolution. 
The pipeline produces results comparable with scientific radiometric high-end thermal cameras in terms of the quality and accuracy of the temperature and CWSI estimations using affordable hardware.
\paragraph*{Declaration of competing interest}
The authors declare no conflict of interest or competing interest.
\paragraph*{Data availability}
\begin{itemize}
    \item The data collected by the UAV and used for \cref{sec:results:real}:
          \begin{enumerate}
              \item The raw IR frames taken by the FLIR \taucamera are available at \url{https://storage.cloud.google.com/end2endtempsr-bkt/realdata/tau2.npz}.
              \item The GT temperature maps collected by the \scientificCamera are available at \url{https://storage.cloud.google.com/end2endtempsr-bkt/realdata/a655sc.npy}.
          \end{enumerate}
    \item The calibration data for the simulator in \cref{sec:methods:nuc} can be found at \url{https://storage.cloud.google.com/end2endtempsr-bkt/CalbrationData.zip}.
    \item The accurate temperature maps used for training and validation in \cref{sec:methods:data} are available upon request from the corresponding author.
\end{itemize}
\paragraph*{Code availability}
The code is available at \url{https://github.com/navotoz/End2EndTemperatureSR/}.
\paragraph*{Funding}
The research was supported by the Israeli Ministry of Agriculture's Kandel Program [grant number 20-12-0030].
\paragraph*{Acknowledgements}
The authors thank Dr. Eitan Goldstein for the data used for the CWSI calculations in this work.
The authors are deeply grateful for the help of Ohaliav Keisar with the UAV data collection.
The authors would also like to thank Moti Barak, Lavi Rosenfeld and Liad Reshef for the design and construction of the environmental chamber.
\paragraph*{CRediT authorship contribution statement}
\textbf{Navot Oz}: Conceptualization, Methodology, Software, Validation, Formal analysis, Investigation, Data curation, Writing - Original Draft, Writing - Review \& Editing, Visualization.
\textbf{Nir Sochen}: Supervision, Writing - Review \& Editing.
\textbf{David Mendlovic}: Supervision, Writing - Review \& Editing.
\textbf{Iftach Klapp}: Conceptualization, Methodology, Resources, Writing - Review \& Editing, Supervision, Project administration.
\clearpage
\bibliographystyle{cas-model2-names}
\bibliography{bibliography}
\newcommand{\sizesAppendixSubfloatsLinePlotsConvergence}{0.25\linewidth}%
\newcommand{\sizesAppendixFontLinePlotsConvergence}{\footnotesize}%
\newcommand{\plotAppendixLineConvergenceMAE}[1]{\addplot[color=blue] table [x=epoch, y=mae, col sep=comma] {#1};}%
\newcommand{\plotAppendixLineConvergenceLR}[1]{\addplot[color=red, mark=none] table [x=epoch, y=lr_sr, col sep=comma] {#1};
    \addlegendentry{SR}
    \addplot[color=black, mark=none] table [x=epoch, y=lr_nuc, col sep=comma] {#1};
    \addlegendentry{NUC}
}%
\newcommand{\appendixPlotAxisMAE}[3]{\begin{tikzpicture}
        \begin{axis}[
                width=#1,
                font=#2,
                xlabel={Epoch},
                ylabel={MAE [$^\circ \text{C}$]},
                grid
            ]
            \plotAppendixLineConvergenceMAE{#3}
        \end{axis}%
    \end{tikzpicture}
}%
\newcommand{\appendixPlotAxisLR}[3]{\begin{tikzpicture}
        \begin{axis}[
                width=#1,
                font=#2,
                xlabel={Epoch},
                ylabel={Learning rate},
                legend style={at={(0.98,0.98)}, anchor=north east},
                grid
            ]
            \plotAppendixLineConvergenceLR{#3}
        \end{axis}
    \end{tikzpicture}
}%
\appendix
\section*{Appendix}
\setcounter{figure}{0}
\renewcommand{\thefigure}{A\arabic{figure}}
\begin{figure*}[!h]
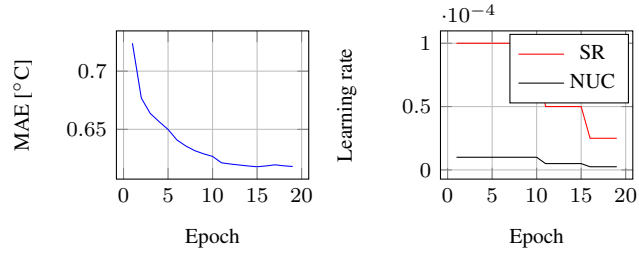
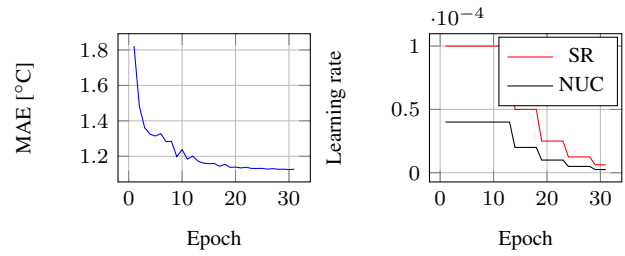
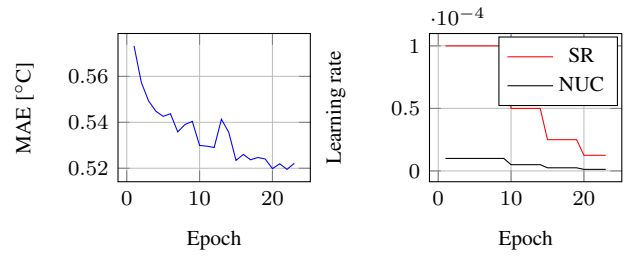
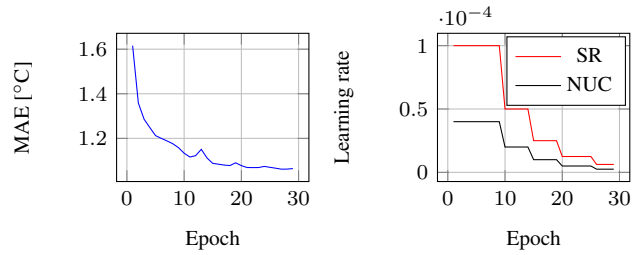

    \centering
    \subfloat[$\times2$, Single-frame nonuniformity correction]{\subfloat{\appendixPlotAxisMAE{\sizesAppendixSubfloatsLinePlotsConvergence}{\sizesAppendixFontLinePlotsConvergence}{figs/appendix/convergence/sr2_single_feature0.csv}}
        \hfill
        \subfloat{\appendixPlotAxisLR{\sizesAppendixSubfloatsLinePlotsConvergence}{\sizesAppendixFontLinePlotsConvergence}{figs/appendix/convergence/sr2_single_feature0.csv}}
        \setcounter{subfigure}{1}
    }\\
    \subfloat[$\times4$, Single-frame nonuniformity correction]{\subfloat{\appendixPlotAxisMAE{\sizesAppendixSubfloatsLinePlotsConvergence}{\sizesAppendixFontLinePlotsConvergence}{figs/appendix/convergence/sr4_single_feature0.csv}}
        \hfill
        \subfloat{\appendixPlotAxisLR{\sizesAppendixSubfloatsLinePlotsConvergence}{\sizesAppendixFontLinePlotsConvergence}{figs/appendix/convergence/sr4_single_feature0.csv}}
        \setcounter{subfigure}{2}
    }\\
    \subfloat[$\times2$, Multiframe nonuniformity correction]{\subfloat{\appendixPlotAxisMAE{\sizesAppendixSubfloatsLinePlotsConvergence}{\sizesAppendixFontLinePlotsConvergence}{figs/appendix/convergence/sr2_multi_feature0.csv}}
        \hfill
        \subfloat{\appendixPlotAxisLR{\sizesAppendixSubfloatsLinePlotsConvergence}{\sizesAppendixFontLinePlotsConvergence}{figs/appendix/convergence/sr2_multi_feature0.csv}}
        \setcounter{subfigure}{3}
    }\\
    \subfloat[$\times4$, Multiframe nonuniformity correction]{\subfloat{\appendixPlotAxisMAE{\sizesAppendixSubfloatsLinePlotsConvergence}{\sizesAppendixFontLinePlotsConvergence}{figs/appendix/convergence/sr4_multi_feature0.csv}}
        \hfill
        \subfloat{\appendixPlotAxisLR{\sizesAppendixSubfloatsLinePlotsConvergence}{\sizesAppendixFontLinePlotsConvergence}{figs/appendix/convergence/sr4_multi_feature0.csv}}
        \setcounter{subfigure}{4}
    }%
    \caption{Training convergence of the mean absolute error (MAE) in $^\circ$C, and learning rate. SR and NUC denote super resolution and nonuniformity correction, respectively.}
    \label{appendix:fig:convergence}
\end{figure*}%

\end{document}